\documentclass[%
 onecolumn,
 amsmath,amssymb,
 aps,
]{revtex4-2}

\usepackage{graphicx}
\usepackage{dcolumn}
\usepackage{bm}
\usepackage{orcidlink} 

\begin{document}

\preprint{APS/123-QED}

\title{Epidemic dynamics in physical-information-social multilayer networks}

\author{Mengshou Wang\textsuperscript{1}}
\author{Liangrong Peng\textsuperscript{2}}%
\author{Baoguo Jia\textsuperscript{3}}%
\author{Liu Hong\textsuperscript{1}}%
\thanks{Corresponding author. hongliu@sysu.edu.cn}

\affiliation{%
	\textsuperscript{1}
	School of Mathematics, Sun Yat-sen University,
	Guangzhou 510275, People’s Republic of China\\
	\textsuperscript{2}
	School of Computer and Data Science, Minjiang University, Fuzhou 350108, People’s Republic of China\\
	\textsuperscript{3} School of Science, Sun Yat-Sen University, Shenzhen 518107, People’s Republic of China
}%



%

\date{\today}

\begin{abstract}
During epidemic outbreaks, information dissemination enhances individual protection, while social institutions influence the transmission through measures like government interventions, media campaigns, and hospital resource allocation. Here we develop a tripartite physical-information-social epidemic model and derive the corresponding kinetic equations in different scales by using the Microscopic Markov Chain Approach and mean-field approximations. The basic reproduction number and epidemic thresholds are explicitly derived by the next generation matrix method. Our results reveal that (1) active information exchange curbs disease transmission, (2) earlier and stronger government responses reduce the epidemic size, and (3) stronger governmental influence on media and hospitals further decreases disease transmission, particularly in hospital nodes. In fixed community structures, groups with frequent physical contact but weak information access (e.g., students) exhibit higher infection rates. For diverse communities, weaker physical layer heterogeneity but stronger information layer heterogeneity (e.g., high internet penetration in rural areas) inhibits epidemic outbreaks. These findings offer valuable insights for epidemic prevention and control strategies.
\end{abstract}

\maketitle


\section{\label{introduction}Introduction}

The explosive spread of epidemic diseases poses significant threats to the economy, society, and public health of every nation in the world, as exemplified by the H1N1 influenza \cite{malikOutliningRecentUpdates2024} and the more recent COVID-19 pandemic \cite{plattoCOVID19AnnouncedPandemic2020}.
Therefore, an in-depth exploration on the dynamic evolution, prevalence, and threshold of epidemics is of significant practical importance, providing a theoretical foundation and practical guidance for both epidemic prevention and control.

Since Bernoulli proposed the first mathematical approach of disease transmission in 1760 \cite{dietzDanielBernoullisEpidemiological2002}, the mathematical modeling of epidemic spread has advanced tremendously, particularly with compartmental models that categorize populations into distinct groups \cite{keelingModelingInfectiousDiseases2008}. In the late 20th century, with the rapid development of network science \cite{albertErrorAttackTolerance2000, newmanNetworksIntroduction2010}, many efforts were also made in the study of epidemic spreading on complex networks \cite{pastor-satorrasEpidemicSpreadingScaleFree2001, pastor-satorrasEpidemicProcessesComplex2015a}. 

The spread of epidemics is often accompanied by the dissemination of disease-related information, which subsequently influences individual behaviors (such as wearing masks and avoiding crowded places) to reduce the probability of infection \cite{haysEpidemicsPandemicsTheir2005,funkEndemicDiseaseAwareness2010}. Consequently, in recent years, the study of coupled disease-information dynamics on complex networks has gained dramatic popularity. Based on the mean-field approach, \cite{zhanCouplingDynamicsEpidemic2018,Samanta2013} indicates that a high disease prevalence rate leads to a slow information decay, which maintains high infection levels while paradoxically suppressing further epidemic spread. Nevertheless, the mean-field approximation fails to consider the network architecture and cannot account for the effects of network heterogeneity on epidemic propagation, thus driving the development of modified mean-field approaches. In the homogeneous mean-field approach, information propagation remains uncoupled from the disease transmission dynamics \cite{Wu2012} , enabling analytical derivation of the basic reproduction number $R_{0}^{d}$ as well as a rigorous proof of its associated global asymptotic stability \cite{Liu2018}. A partial effectiveness approach combining the effective degree theory for one network layer and the heterogeneous mean-field approximation for another was proposed, demonstrating both preserved accuracy in epidemic prediction and reduced dimensionality of governing equations in computations \cite{Zhou2019}. The Markov chain-based approach (MMCA) was developed to investigate the coupled spreading dynamics of awareness and epidemics in multilayer networks, revealing that the epidemic threshold is fundamentally determined by the largest eigenvalue of the physical layer's adjacency matrix modified by information diffusion \cite{granellDynamicalInterplayAwareness2013,Granell2014}. The corresponding continuous-time MMCA method yields the associated basic reproduction number \cite{Huang2018}. Furthermore, regarding the continuous-time and discrete-time mean-field equations for epidemic spreading on complex networks, remarkable differences in disease prevalence emerge near the epidemic threshold \cite{Silva2024}.

Early coupled disease-information multilayer network models primarily adopted the SIS-UAU framework, with the physical layer following susceptible-infected-susceptible dynamics and the information layer modeling unaware-aware-unaware transitions \cite{granellDynamicalInterplayAwareness2013,Granell2014,Pan2018}. More sophisticated models capturing complex mechanisms have been extensively investigated, including SIR-UAU \cite{Zheng2018,Xia2019,Wang2019},SEIR-UAU\cite{Ma2022}, SIRD-UAU\cite{Zhu2019}, and SACIR-UAU \cite{He2021} variants. Extended frameworks incorporating immunization behaviors or rumor-related influences like the UAU-DKD-SIQS model have also been examined \cite{Yu2025,Yu2025a}. Recent studies highlight the role of information and activity dynamics in epidemic control. \cite{Chen2023} found that optimal coupling between information and activity preferences effectively suppresses epidemics. However, \cite{Chen2025} introduced resource factors, showing that excessive information flow intensifies resource competition, leading to inefficient allocation and worsened outbreaks.

However, few studies have incorporated governmental nodes influencing both physical and information layers, with these entities typically assumed to maintain constant impacts \cite{Xia2019,Ma2022,Li2024}. Therefore, in the current study, we propose a governmental node mechanism that dynamically adjusts intervention strategies based on real-time surveillance of  epidemic prevalence, directly modulating the transmission probabilities while indirectly influencing the hospital treatment capacities and media-driven information dissemination. Our results demonstrate that earlier and stronger government responses lead to a reduced epidemic size. Similarly, increased government influence on the media and hospitals, particularly the latter, contributes to a more effective epidemic control. We found that in fixed community structures, groups with more physical contacts and less information exchange, such as primary and secondary school students, exhibit higher infection rates. For different community structures, weaker heterogeneity in the physical layer and stronger heterogeneity in the information layer result in higher epidemic thresholds, making outbreaks less likely, as observed in rural areas with high internet penetration rates. We also simulated networks with community structures, finding epidemics spread faster to adjacent communities.


\section{Basic Model}

\begin{figure*}
	\centering
	\includegraphics[width=1.0\linewidth]{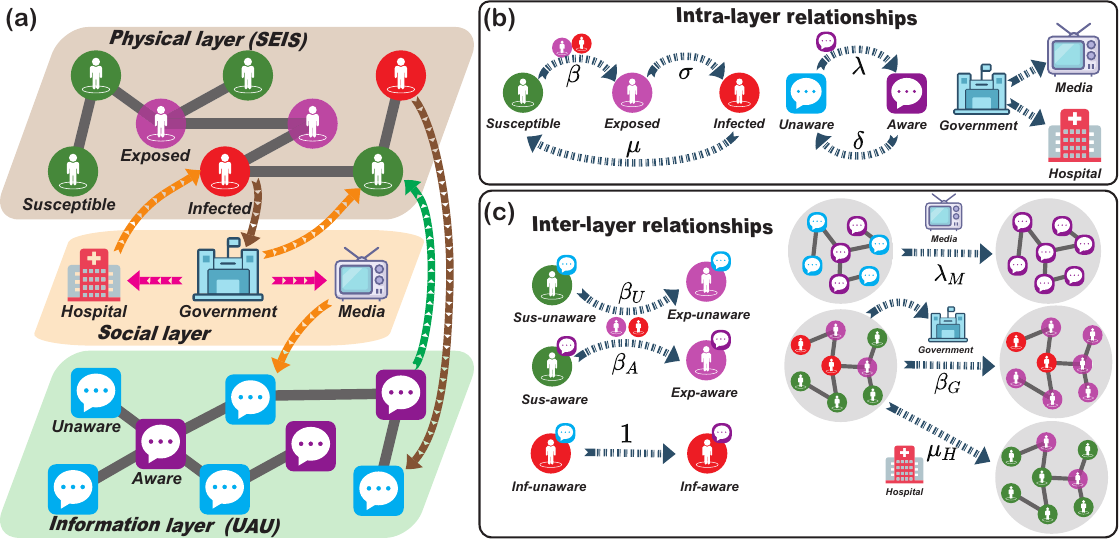}
	\caption{Physical-information-social triadic space network model for epidemic spreading. (a) illustrates the SEIS model in the physical layer, the UAU model in the information layer, and the government, media \& hospital in the social layer. (b) depicts the intra-layer state transition mechanisms across the three distinct layers. (c) demonstrates the inter-layer interaction mechanisms.}
	\label{fig1}
\end{figure*}

With respect to a two-layer multiplex networks framework, that has been used to examine the interaction between physical and information spaces in epidemic spread \cite{granellDynamicalInterplayAwareness2013}, we introduce a social space as a third layer to capture the influence of social institutions on epidemic transmission, providing a more realistic model for the effective social interventions in disease control.

As illustrated in Fig.~\ref{fig1}(a), there is a one-to-one mapping between the nodes in the physical layer (P-layer) and in the information layer (I-layer), indicating that \(N\) individuals exist in both layers. In the social layer (S-layer), we consider three nodes representing social institutions: the government, the media and hospital departments, which can influence all nodes in the I-layer and the P-layer, respectively. For simplicity, the connections considered here are undirected and unweighted.

As shown in Fig.~\ref{fig1}(b), in the P-layer, we use the susceptible-exposed-infected-susceptible (SEIS) model to describe the epidemic transmission. The exposed state (E) represents infected individuals without clinical symptoms, while the infected state  (I) represents symptomatic individuals. Both exposed (E) and infectious (I) nodes can infect neighboring susceptible (S) nodes with a probability of $\beta$, converting them into the exposed (E) state, where $\beta$ is referred to as the \textit{infection probability}. The exposed (E) state transits to the infectious (I) state with a probability of $\sigma$, referred to as the \textit{exposed-to-infectious probability}, while the infectious (I) state transits back to the susceptible (S) state with a probability of $\mu$, termed the \textit{recovery probability}.

In the I-layer, we employ the unaware-aware-unaware (UAU) model to describe the information dissemination. Individuals in the unaware state (U) lack information about the epidemic and take no protective measures, while those in the aware state (A) are informed and adopt protective actions, such as wearing masks. Individuals in state U transit to state A with a probability $\lambda$ when influenced by neighboring  individuals in state A, termed the \textit{awareness probability}. Conversely,  individuals in state A revert to state U with a probability \(\delta\) due to the forgetting mechanism, termed the \textit{forgetting probability}.

\begin{figure}
	\centering
	\includegraphics[width=0.8\linewidth]{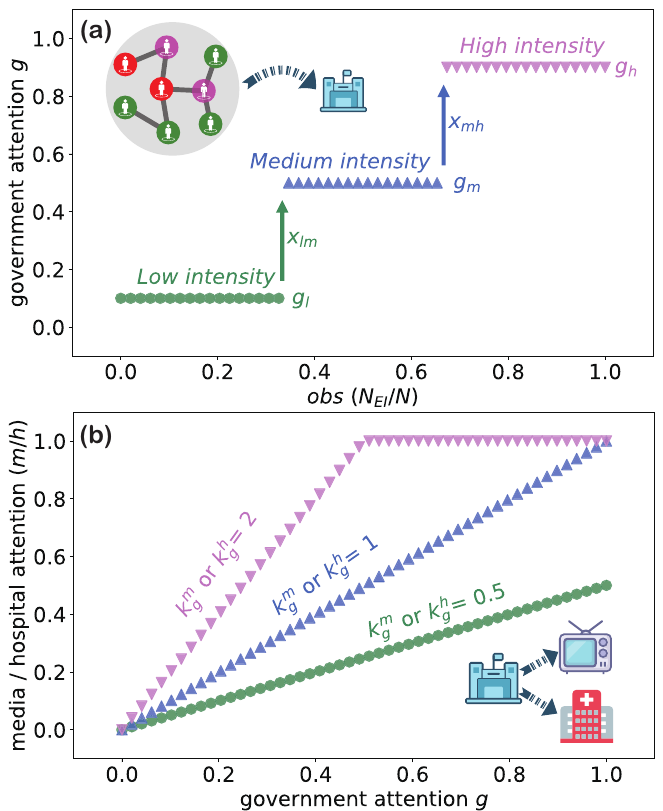}
	\caption{Plots of the response functions $f_{obs}^g$, $f_g^m$, and $f_g^h$, representing the relationships between the observed values and the government attention, and between the government attention and media/hospital attention, respectively.}
	\label{fig2}
\end{figure}

In the S-layer, we focus on government, media and hospital departments. Let $g$ denote the level of attention paid by the government to the epidemic, where $g \in [0,1]$, termed the \textit{government attention}. Here, $g = 0$ indicates that the government is completely indifferent to the epidemic, whereas $g = 1$ signifies that the government is fully attentive to the epidemic. The attention levels of media and hospital nodes are similarly represented by $m$ and $h$, respectively, termed the \textit{media attention} and the \textit{hospital attention}. We hypothesize that the media attention to the epidemic is influenced by the government attention via a mapping function \( f_g^m: [0,1] \rightarrow [0,1] \), such that \( m = f_g^m(g) \). As illustrated in Fig.~\ref{fig2}(b), a specific formulation could be a truncated linear function:
\[
f_g^m(x) = \text{clip}(k x, 0, 1),
\]
where \( k=k_g^m \ge 0 \) reflects the degree of the government attention influence on the media, termed the \textit{media sensitivity}, and \(\text{clip}(x, a, b)\) denotes the clipping function that outputs \( a \) if \( x < a \), \( b \) if \( x > b \), and \( x \) otherwise. The hospital attention to the epidemic is influenced by the government attention in a similar manner, modeled by a mapping function $f_g^h: [0,1] \rightarrow [0,1]$, such that \(h = f_g^h(g) \). Again, the truncated linear function can be adopted, where $k_g^h  \geq 0$ reflects the degree of the government attention influence on the hospital, termed the \textit{hospital sensitivity}.

Having established the intra-layer relationships within each layer, we now outline the inter-layer relationships in Fig.~\ref{fig1}(c): 
\begin{enumerate}
	\item[1)] I-layer $\Leftrightarrow$ P-layer: the co-located node state SA (where a node belongs to both S and A) reduces the  infection probability \(\beta_A\), formulated as:
	$$
	\beta_A = \beta \gamma_A,
	$$
	where \(\gamma_A \in [0,1]\) represents the transmission rate discount factor due to protective measures.
	Conversely, the SU state (a node in both S and U) retains the baseline infection probability:
	\[
	\beta_U = \beta.
	\]
	The node IU transits to IA with probability 1, under the assumption that symptomatic infected individuals invariably become aware of the disease.

	\item[2)] S-layer $\Leftrightarrow$ P-layer: the infection level within the population influences the government attention to the epidemic. 
	We define the observational metrics $obs$ for the infection severity:
			$$obs = \frac{N_{EI}}{N},$$
			where $N_{EI}$ is the total number of nodes in states E and I.
	
	The government attention \( g \) is determined by a mapping function \( f_{obs}^g \), such that \( g = f_{obs}^g(obs) \). 
	Based on the value of $obs$, the government attention is divided into three stages: \textit{low intensity}, \textit{medium intensity}, and \textit{high intensity}, corresponding to attention values $g_l$, $g_m$, and $g_h$, respectively. The boundary between low and medium intensity is defined by $x_{lm}$, termed the \textit{low-medium boundary}, while the boundary between medium and high intensity is defined by $x_{mh}$, termed the \textit{medium-high boundary}, as illustrated in Fig.~\ref{fig2}(a). The expression for the attention level is defined as 
	$$
	f_{obs}^g(x) =
	\begin{cases}
		g_l, &x \in [0, x_{lm}),\\
		g_m, &x \in [x_{lm}, x_{mh}),\\
		g_h, &x \in [x_{mh}, 1].\\
	\end{cases}
	$$
	Government interventions in response to the epidemic (e.g., closing entertainment venues) reduce the infection probability. Specifically, let
	\[
	\beta_G = \beta(1 - g)
	\]
	denote the government-driven infection probability. If susceptible individuals are also influenced by awareness, then the infection probability becomes
	\[
	\beta_{GA} = \beta(1 - g)\gamma_{A}.
	\]
	
	A higher level of hospital attention to the epidemic leads to a greater allocation of medical resources toward patients under treatment, thereby increasing the recovery probability $\mu^{\prime}$. We set $\mu^{\prime} = h$, where $h$ represents the hospital attention to the epidemic. Under the influence of hospital intervention, the hospital-driven recovery probability is given by:
	\[
	\mu_{H} = 1- (1-\mu)(1-\mu^{\prime}) = \mu + \mu^{\prime} - \mu \mu^{\prime}.
	\]
	
	\item[3)] I-layer $\Leftrightarrow$ S-layer: the media coverage of the epidemic enhances the public awareness, thereby increasing the preventive vigilance. We model this behavioral adaptation by introducing a media-driven transition rate \(\lambda_M\), representing the probability that an unaware individual in the I-layer becomes aware due to the media exposure. Formally, we define:
	\[
	\lambda_M = m,
	\]
	where the media attention parameter \(m \in [0,1]\) scales the strength of media influence (a higher \(m\) corresponds to more intensive coverage and thus a greater behavioral shift).
	
\end{enumerate}
\begin{table}
	\caption{Description of symbols}
	\label{Table1}
	\begin{ruledtabular}
		\begin{tabular}{c c } 
			Symbol  & Description  \\ 
			\hline
			P-layer &  \\
			$\beta$ & infection probability (S $\xrightarrow{\rm{E/I}}$ E) \\
			$\sigma$ & exposed-to-infectious probability (E $\rightarrow$ I)   \\ 
			$\mu$ &  recovery probability (I $\rightarrow$ S)	 \\ 
			$(a_{ij})$ & adjacency matrix of the P-layer\\
			\hline
			I-layer & \\
			$\lambda$ & awareness probability (U $\xrightarrow{\rm{A}}$ A)	 \\ 
			$\delta$ &  forgetting probability (A $\rightarrow$ U)	 \\ 
			$(b_{ij})$ & adjacency matrix of the I-layer\\
			\hline
			S-layer & \\
			$g$ & government attention \\
			$m$ & media attention \\
			$h$ & hospital attention\\
			\hline
			$\beta_A$ & infection probability under aware	 \\ 
			$\beta_U$ & infection probability under unaware	 \\ 
			$\beta_G$ & government-driven infection probability \\
			$\beta_{GA}$ & government-driven infection probability under aware	 \\
			$\lambda_{M}$ & media-driven awareness transition probability  \\
			$\mu_{H}$ & hospital-driven recovery probability \\
			$obs$ & fraction of infected population\\
			\hline
			$g_l,g_m,g_h$ & low, medium, high intensity attention\\
			$x_{lm},x_{mh}$ &low-medium, medium-high boundaries\\
			$k_{g}^{m}$ & media sensitivity\\
			$k_{g}^{h}$ & hospital sensitivity\\
		\end{tabular}
	\end{ruledtabular}
\end{table}
In this model, each individual exists in one of five states: SU, SA, EU, EA, IA. The IU states are not included, as we assume that individuals in I are always in A. Table~\ref{Table1} summarizes the corresponding symbolic representations.

\section{MMCA approach}
\label{sec3}
\begin{figure*}
	\centering
	\includegraphics[width=0.9\linewidth]{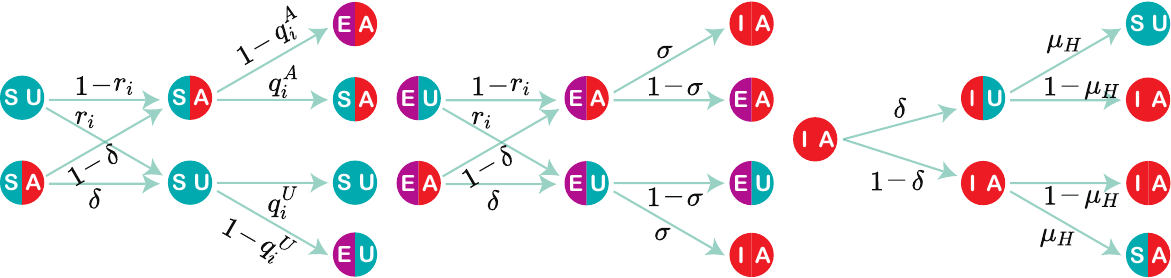}
	\caption{Markov state transition tree}
	\label{fig3}
\end{figure*}
We define \( A = (a_{ij}) \) (resp. \( B = (b_{ij}) \)) as the adjacency matrix of the P-layer (resp. I-layer), where \( a_{ij} = 1 \) (resp.  \( b_{ij} = 1 \)) if node \( i \) and node \( j \) are connected, and \( a_{ij} = 0 \) (resp.  \( b_{ij} = 0 \)) otherwise. At time \( t \), each individual \( i \) (for \( i = 1, \dots, N \)) can be in one of five states, with corresponding probabilities denoted as \( p_{i}^{SU}(t) \), \( p_{i}^{SA}(t) \), \( p_{i}^{EU}(t) \), \( p_{i}^{EA}(t) \), and \( p_{i}^{IA}(t) \). Let \( p^{U}(t) = \sum_{i} \left(p_{i}^{SU}(t) + p_{i}^{EU}(t)\right)/N \) and \( p^{I}(t) = \sum_{i} p_{i}^{IA}(t)/N \).

For each possible state of a node at time \( t \), it will transit to the corresponding states at time \( t+1 \). In Fig.~\ref{fig3}, we present the Markov state transition tree for node \( i \) (where \( i=1,...,N \)).
In the I-layer, we define \( r_{i}(t) \) as the probability that node \( i \) in the U will not be informed by any neighbors. In the P-layer, we define \( q_{i}^{U}(t) \) and \( q_{i}^{A}(t) \) as the probabilities that node \( i \) in the SU and SA, respectively, will not be infected by any neighbors at time \( t \). From the above definitions, we can obtain:
\begin{equation}\label{eq3}
	\begin{aligned}
		r_{i}(t) =& (1-\lambda_{M}(t))\prod_{j}\left[1-b_{ji}p_{j}^{A}(t)\lambda\right],\\
		q_{i}^{U}(t) =& \prod_{j}\left\{1-a_{ji}p_{j}^{EI}(t)\beta_G(t)\right\},\\
		q_{i}^{A}(t) =&\prod_{j}\left\{1-a_{ji}p_{j}^{EI}(t)\beta_{GA}(t)\right\},
	\end{aligned}
\end{equation}
where $p_{i}^{A}(t) = p_{i}^{SA}(t)+p_{i}^{EA}(t)+p_{i}^{IA}(t)$ and $p_{i}^{EI}(t) = p_{i}^{EU}(t)+p_{i}^{EA}(t)+p_{i}^{IA}(t).$ 
The observation function $obs(t)$ is defined as:
$$obs(t) = \frac{\sum_{i} p^{EI}_i(t)}{N}.$$
Note this defition agrees with the previous one $obs=N_{EI}/N$ in the limit of infinitely large population number.

Based on $obs(t)$, the transmission rate $\beta_{G}(t)$ is defined as:
$$\beta_{G}(t) = \beta \left(1 - f_{obs}^g(obs(t))\right).$$
The adjusted transmission rate $\beta_{GA}(t)$ is given by:
$$
\beta_{GA}(t) = \beta_{G}(t) \gamma_{A}.
$$
The mortality rate $\lambda_{M}(t)$ is defined as:
$$
\lambda_{M}(t) = f_g^m \circ f_{obs}^g(obs(t)),
$$
and the recovery rate $\mu_{H}(t)$ is expressed as:
$$
\mu_{H}(t) = \mu + f_g^h \circ f_{obs}^g(obs(t)) - \mu \left( f_g^h \circ f_{obs}^g(obs(t)) \right).
$$

Referring the Markov state transition tree, we can derive the corresponding MMCA equations: 
\begin{equation}\label{MMCA}
	\begin{aligned}
		p_{i}^{SU}(t+1) =& r_{i}(t)q_{i}^{U}(t)p_{i}^{SU}(t)+\delta q_{i}^{U}(t)p_{i}^{SA}(t) \\&+ \delta \mu_{H} p_{i}^{IA}(t),\\
		p_{i}^{SA}(t+1) =& (1-r_{i}(t))q_{i}^{A}(t)p_{i}^{SU}(t)\\&+(1-\delta) q_{i}^{A}(t)p_{i}^{SA}(t)\\&
		+ (1-\delta) \mu_{H} p_{i}^{IA}(t), \\
		p_{i}^{EU}(t+1) =& r_{i}(t)(1-q_{i}^{U}(t))p_{i}^{SU}(t)\\&+\delta (1-q_{i}^{U}(t))p_{i}^{SA}(t)
		\\&+ r_{i}(t)(1-\sigma)p_{i}^{EU}(t) \\&+ \delta(1-\sigma)p_{i}^{EA}(t),\\
		p_{i}^{EA}(t+1) =& (1-r_{i}(t))(1-q_{i}^{A}(t))p_{i}^{SU}(t)\\&+(1-\delta) (1-q_{i}^{A}(t))p_{i}^{SA}(t) \\&
		+ (1-r_{i}(t))(1-\sigma)p_{i}^{EU}(t)\\& + (1-\delta)(1-\sigma)p_{i}^{EA}(t),\\
		p_{i}^{IA}(t+1) =& \sigma p_{i}^{EA}(t)+\sigma p_{i}^{EU}(t)+(1-\mu_{H}) p_{i}^{IA}(t).
	\end{aligned}
\end{equation}

The above description of epidemic dynamics is formulated at the individual level, hence we refer to it as the discrete microscopic mean-field equations. In Appendix \ref{CMMCA}, by considering the time step $\Delta t$ goes to zero $\Delta t\rightarrow 0$ and reorganizing Eq.~\eqref{MMCA}, we derive the continuous microscopic mean-field equations, which are also termed the quenched mean-field equations.

Assuming nodes with identical degrees in both physical and information layers are statistically equivalent, we derive the mesoscopic mean-field equations in Appendix \ref{degreeMMCA}. Further, under the assumption that all nodes are statistically equivalent, the macroscopic mean-field equations are established in Appendix \ref{MacroscopicMMCA}. These form the basis for analyzing epidemic thresholds across three distinct scales in the following section.

\section{Epidemic threshold}
\label{sec4}
In the theoretical study of epidemics, the epidemic threshold plays a crucial role, as it determines whether an outbreak will occur or not. Building on the mean-field equations for epidemic transmission dynamics established in the previous section, we now proceed to analyze the epidemic threshold at three different scales. We formally present the corresponding basic reproduction number:
\begin{equation}\label{R0}
	R_0 = \beta R^{G}_{0} R^{H}_{0} R^{M}_{0},
\end{equation}
where $R^{G}_{0}$, $R^{H}_{0}$, and $R^{M}_{0}$ represent the impacts of the government, hospitals, and the media, respectively, on epidemic transmission. It is noteworthy that both discrete and continuous formulations yield identical results at the same level.
We observe that $R^{G}_{0}$ and $R^{H}_{0}$ share the same form across the three levels, given by:
$$
R^{G}_{0} = 1 - g_l, \quad R^{H}_{0} = \frac{1}{\sigma} + \frac{1}{\mu_{H}^{0}},
$$
where $\mu_{H}^{0} = \mu+k_g^h g_l-\mu k_g^h g_l.$
However, $R^{M}_{0}$ exhibits different representations depending on the level. Detailed derivations are provided in Appendixes \ref{proofs in MMCA}, \ref{R0_QMF}, \ref{degreebasedMMCA}  \cite{VanDenDriessche2002}.

Since the epidemic does not emerge when the basic reproduction number $R_0 < 1$, we can derive the epidemic threshold as follows:
\begin{equation}
	\beta_c = \frac{1}{ R^{G}_{0} R^{H}_{0} R^{M}_{0}}.
\end{equation}

\subsection{Microscopic level}
%
%
In Appendix \ref{R0_QMF}, analysis at the individual level reveals the influence of media on the basic reproduction number $R_0$, expressed as 
\begin{equation}\label{R0M}
	R_0^M = \Lambda_{\max}(H),
\end{equation}
where $\Lambda_{\max}(H)$ denotes the dominant eigenvalue of matrix $H$, encapsulating the aggregate impact of media dissemination on the disease transmission dynamics. 
The elements of matrix \( H \) read 
\begin{equation}\label{H_element}
	h_{ij} = \left(1-(1-\gamma_A)p_{i}^{A}\right)a_{ji} ,
\end{equation}
and $p_{i}^{A}$ represents the steady-state probability of the $i$-th node in the single-layer I-layer, which is obtained through iterative computations using relations
\begin{equation}\label{eqpA}
	p_{i}^{A} = (1-r_{i})+(r_{i}-\delta) p_{i}^{A}
\end{equation}
and
\begin{equation}\label{eqr}
	r_{i}=\left[1-k_g^m g_l\right]\prod_{j}\left[1-b_{ji}p_{j}^{A}\lambda\right].
\end{equation}

Under two specific conditions, the parameter $R^M_0$  reduces to a simplified form.
\begin{itemize}
\item If we approximate $p_{i}^{A}$ using 
\begin{equation}\label{eq13}
	p^{A} = \frac{1}{N} \sum_{i} p_{i}^{A}
\end{equation}
\cite{Chang2021}, we find that 
\begin{equation}\label{R_0M_approx1}
	R_0^M = \left(1-(1-\gamma_A)p^{A}\right)\Lambda_{\max}(A),
\end{equation}
where \(\Lambda_{\max}(A)\) denotes the largest eigenvalue of the P-layer adjacency matrix.
Similarly, here we iteratively solve for $p_i^{A}$ using Eq.\eqref{eqpA} and Eq.\eqref{eqr}, thereby obtaining $p^{A}$.
\item  When \(\lambda = 0\) (i.e., in an information society where disease transmission relies solely on the media dissemination), we find that 
\begin{equation}\label{R_0M_approx2}
	R_0^M = \frac{\gamma_{A}+\delta_m}{1+\delta_m}\Lambda_{\max}(A),
\end{equation}
where $ \delta_m = \delta/(k_g^m g_l) $.
\end{itemize}

\subsection{Mesoscopic level}

Assuming nodes with identical degrees in both physical and information layers are statistically equivalent, $R_0^M$ yields results analogous to those at the microscopic level, i.e.,
\begin{equation}\label{R0M1}
	R_0^M = \Lambda_{\max}(H),
\end{equation}
where the elements of the matrix $H$ are 
\begin{equation}\label{H_element2}
	h_{\mathbf{k^{\prime}}\mathbf{k}} = \left(1-(1-\gamma_A)p_{\mathbf{k}}^{A}\right)k_1P_1(\mathbf{k^{\prime}}|\mathbf{k}),
\end{equation}
and the matrix is a square matrix of size $k_1^{\max} k_2^{\max}$.
Here, 
\begin{equation}\label{pA1}
	p_{\mathbf{k}}^A = \frac{1 - \left[1 - k_g^m g_l\right] \left[1 - \lambda \Phi_{\mathbf{k}}\right]^{k_2}}{\delta + 1 - \left[1 - k_g^m g_l\right] \left[1 - \lambda \Phi_{\mathbf{k}}\right]^{k_2}},
\end{equation}
and $\Phi$ can be iteratively obtained through the equation
\begin{equation}\label{phi1}
	\Phi_{\mathbf{k}} = \sum_{\mathbf{k^{\prime}}} P_2(\mathbf{k^{\prime}} | \mathbf{k}) \frac{1 - \left[1 - k_g^m g_l\right] \left[1 - \lambda \Phi_{\mathbf{k^{\prime}}}\right]^{k_2^{\prime}}}{\delta + 1 - \left[1 - k_g^m g_l\right] \left[1 - \lambda \Phi_{\mathbf{k^{\prime}}}\right]^{k_2^{\prime}}}.
\end{equation}

\begin{itemize}
\item Particularly, when the P-layer and I-layer networks are independent and both uncorrelated, we have
\begin{equation}\label{R0M2}
	R_0^M =\left(1-(1-\gamma_A)p^{A}\right)\frac{\langle k_1^2 \rangle}{\langle k_1 \rangle},
\end{equation}
Here, the probability $p^{A}$ is expressed as 
\begin{equation}\label{pA2}
	p^A = \sum_{k_2} P_{I}(k_2) \frac{1 - \left[1 - k_g^m g_l\right] \left[1 - \lambda \Phi\right]^{k_2}}{\delta + 1 - \left[1 - k_g^m g_l\right] \left[1 - \lambda \Phi\right]^{k_2}},
\end{equation}
and the quantity $\Phi$ can be iteratively computed using the equation
\begin{equation}\label{phi2}
	\Phi =\frac{1}{\langle k_2 \rangle} \sum_{k_2}k_2 P_{I}(k_2)\frac{1 - \left[1 - k_g^m g_l\right] \left[1 - \lambda \Phi\right]^{k_2}}{\delta + 1 - \left[1 - k_g^m g_l\right] \left[1 - \lambda \Phi\right]^{k_2}}.
\end{equation}
Note that when $k_{g}^{m}g_l = 0$, $p^A$ admits an approximate expression, as discussed in \cite{PastorSatorras2001}.
\end{itemize}

\subsection{Macroscopic level} 
Under the assumption that all nodes are statistically equivalent, meaning the network heterogeneity is completely disregarded, $R_0^M$ can be simplified to a more concise expression,
\begin{equation}\label{R0M3}
	R_0^M =\left(1-(1-\gamma_A)p^{A}\right) \langle k_1 \rangle.
\end{equation}
It is noted that the above expression can also be derived by assuming $\langle k_{1}^2 \rangle = \langle k_{1} \rangle^2$ in conjunction with Eq.\eqref{R0M2}.
Here, $p^A$ is the root of the equation $f(x) = 0$ on the interval $[0, 1]$, where 
\begin{equation}\label{pA3}
	f(x) = x^2 + \frac{k_{g}^{m}g_l + \delta - 2\lambda\langle k_2 \rangle}{2\lambda\langle k_2 \rangle} x - \frac{k_{g}^{m}g_l}{2\lambda\langle k_2 \rangle}.
\end{equation}
Since $- \frac{k_{g}^{m}g_l}{2\lambda\langle k_2 \rangle} < 0$, the positive root is selected.

\section{Result and discussion}
For the proposed physical-information-social multilayer network framework for the epidemic spreading, we set the total number of individuals in the network to be 5000, with the uncorrelated configuration model (UCM) for both physical and information layers, whose degree distribution follows $P(k) \sim k^{\frac{5}{2}},$ and a maximum degree cutoff at $N^{\frac{1}{2}}$ is implemented to better reflect realistic network structures. Specifically, the connections between the information and physical layers are independent, with the information layer containing more edges than the physical layer. If all nodes in the network are susceptible and unaware, obviously no epidemic would emerge. Thus to simulate the procedure of epidemic spreading, we randomly select 5\% of P-layer nodes and set their status to be EU. Over time, the EU status transits to IA, facilitating further information and epidemic propagation, and raising responses from the government, hospitals, and the media in the social layer. After a sufficiently long period, the system reaches a steady state at the macroscopic level, where the epidemic size and awareness size are averaged during this phase. 

\subsection{Numerical verification}
To investigate the effectiveness of the MMCA within the theoretical framework presented in Sec.~\ref{sec3}, we employ the Monte Carlo (MC) simulations. Additionally, to assess the accuracy of the theoretical epidemic threshold outlined in Sec.~\ref{sec4}, we utilize a quasi-stationary (QS) analysis with reflective boundaries \cite{Sander2016}. This approach prevents the misjudgment of sub-threshold epidemics due to the system-size-induced extinction. All source codes are publicly available at \url{https://github.com/WangMengshou/PIS-multilayer}.

Here, we define the s-th moment of the QS density of exposed and infected nodes as 
$$[\rho^s] = \frac{1}{N^s} \sum\limits_{n=1}^N n^s \overline{P}_n,$$
where $\overline{P}_n$ represents the QS probability of the system having $n$ infected nodes, computed after a relaxation time $t_r$ during an averaging time $t_{av}$. Furthermore, we adopt the dynamical susceptibility \cite{Ferreira2012}, 
$$[\chi] = N \frac{[\rho^2] - [\rho]^2}{[\rho]},$$ 
which exhibits a pronounced peak at the epidemic threshold for networks without outliers, as illustrated in  Fig~\ref{fig5}(a) \cite{Mata2015}. $[\rho]$ represents precisely the epidemic size $p^{EI}$.

\begin{figure}
	\centering
	\includegraphics[width=0.9\linewidth]{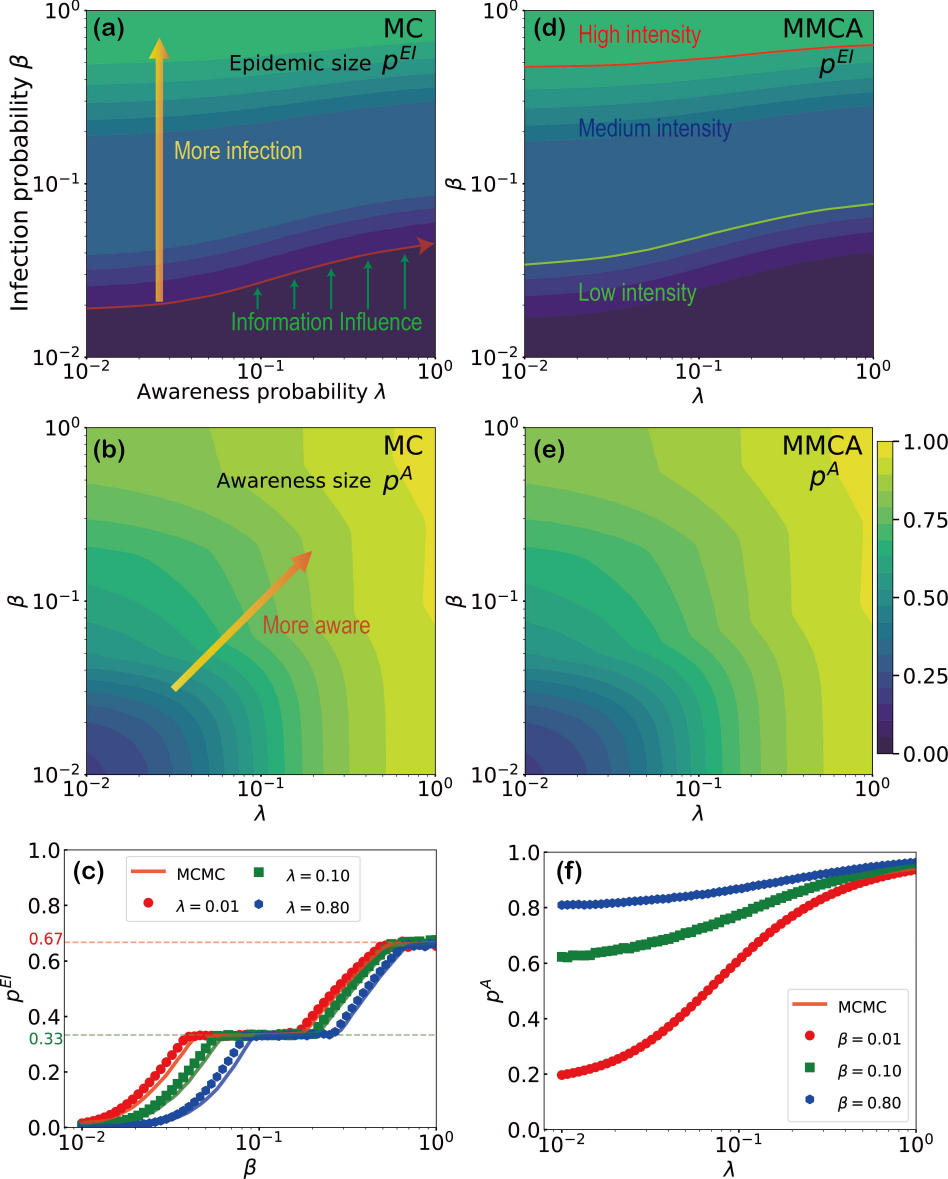}
	\caption{Comparison of the epidemic size $p^{EI}$ (including both exposed and infected individuals) and awareness size $p^A$ between MC numerical simulations and MMCA theoretical analysis. Phase diagrams of  $p^{EI}$ and $p^A$ as functions of the infection probability $\beta$ and awareness probability $\lambda$ are shown in (a) and (b) for MC simulations, and in (d) and (e) for MMCA results. Trajectories of the epidemic size as a function of $\beta$ are depicted in (c), while trajectories of the awareness size as a function of $\lambda$ are shown in (f). Other parameters are fixed at $\sigma = 0.5$, $\mu = 0.1$, $\delta = 0.3$, $\gamma_A = 0.3$, $x_{lm} = 1/3$, $x_{mh} = 2/3$, $g_l = 0.1$, $g_m = 0.5$, $g_h = 0.9$, $k_{gm} = k_{gh} = 0.5$.}
	\label{fig4}
\end{figure}

\begin{figure}
	\centering
	\includegraphics[width=0.7\linewidth]{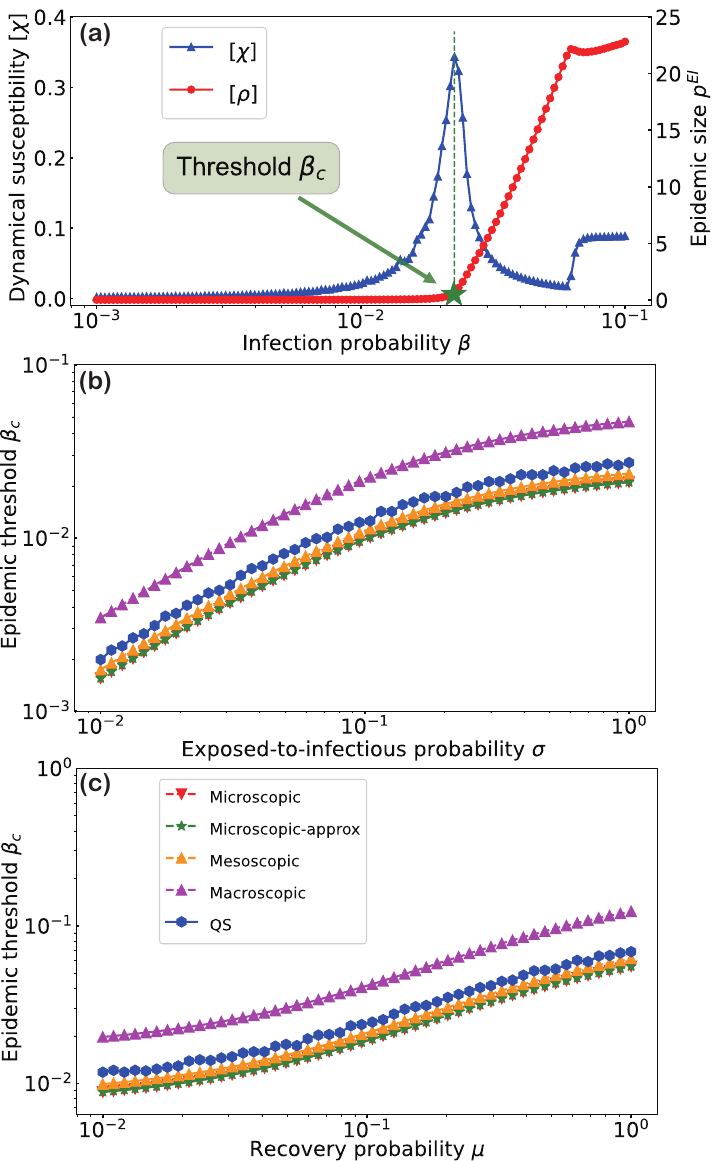}
	\caption{Comparison of epidemic thresholds between quasi-static numerical simulations and theoretical predictions. 
		(a) shows the variation of dynamical susceptibility (blue) and quasi-static epidemic size (red) with respect to the infection probability $\beta$. Theoretical thresholds (four variants) are compared with quasi-static simulation thresholds for different values of the exposed-to-infectious probability $\sigma$ in (b) and the recovery probability $\mu$ in (c). Other parameters are kept consistent with those in Fig.~\ref{fig4}.}
	\label{fig5}
\end{figure}

\begin{figure*}
	\centering
	\includegraphics[width=0.8\linewidth]{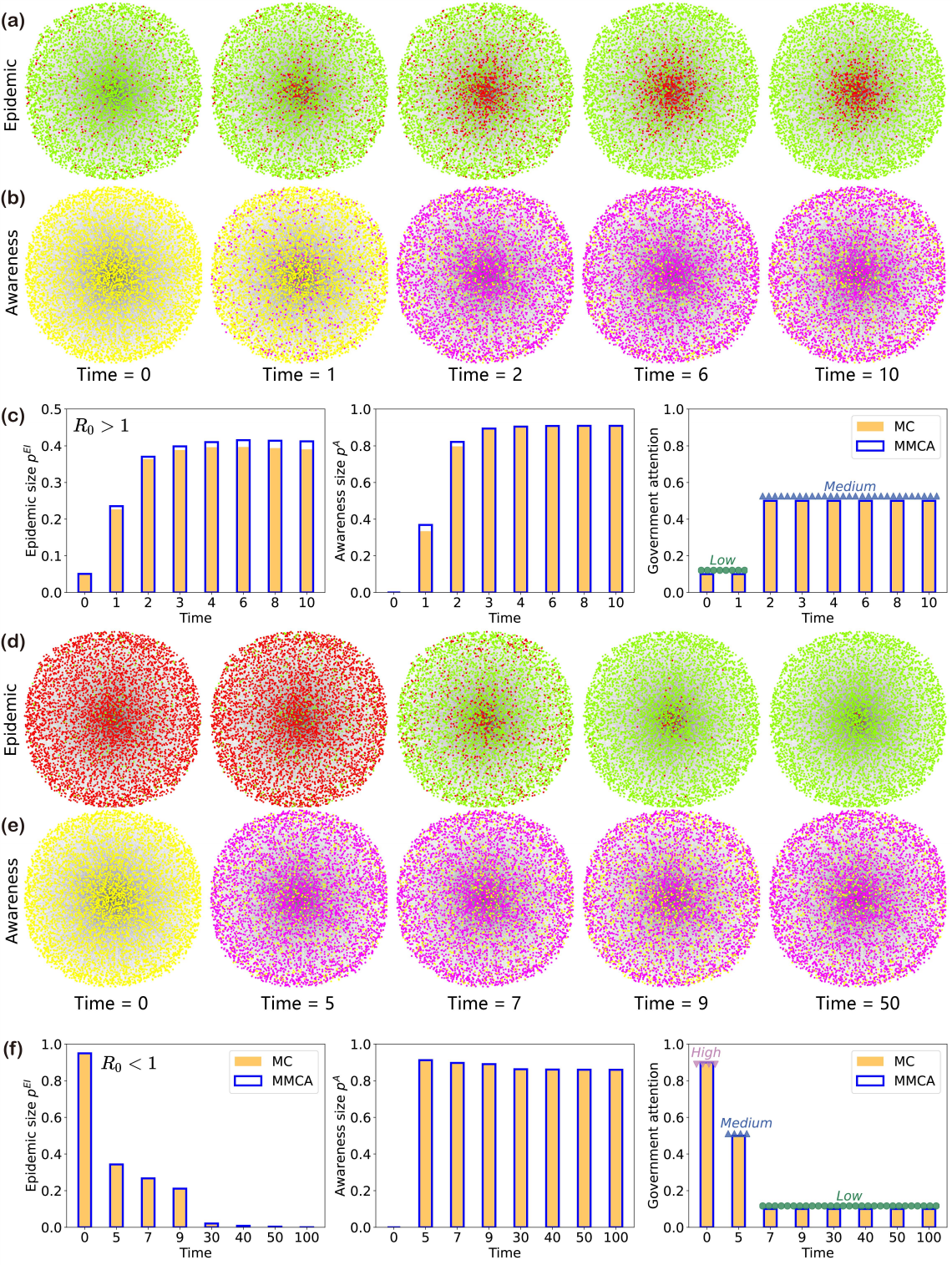}
	\caption{Propagation of the infectious disease on the network. In panels (a) and (b), the spreading of epidemic and awareness are depicted at time points 0, 1, 2, 6, and 10. In (a), the green dots denote the susceptible state (S) and the red dots signify the exposed or infected states (E/I), while in (b), the yellow dots represent the unaware state (U) and the purple dots indicate the aware state (A). In (c), the temporal evolution of the epidemic size (left) and the awareness size (center) is compared using MMCA and MC methods, while the right side illustrates the variation in the government attention over time. Panels (a,b,c) present the results for the case where the basic reproduction number $R_0 > 1$, while panels (d, e, f) correspond to the scenario with $R_0 < 1$. Both physical and information layers are constructed as Barabási-Albert (BA) networks, with parameters $\beta = 0.3$(a,b,c), $\beta = 0.01$(d,e,f) and $\lambda = 0.5$, while other parameters remain consistent with those in Fig.~\ref{fig4}.}
	\label{fig9}
\end{figure*}

To validate the accuracy of the MMCA, MC simulations and calculations using Eq.\eqref{MMCA} were performed for different infection probability $\beta$ and awareness probability $\lambda$, yielding the stationary state epidemic size $p^{EI}$ and awareness size $p^{A}$. A comparison of Fig.~\ref{fig4}(a)(b) and (c)(d) shows that the MMCA accurately captures the phase diagrams of $p^{EI}$ and $p^{A}$. Additionally, Fig.~\ref{fig4}(c) and Fig.~\ref{fig4}(d) provide a detailed illustration of how the MMCA method closely approximates the MC simulations across varying $\beta$ and $\lambda$.

Based on the above discussion, the calculation of $R_0^M$ requires the computation of $p^A$ (or $p^A_i$) at multiple levels. For this purpose, $p^A_i$ is iteratively derived using Eq.\eqref{eqpA} and \eqref{eqr}, while $p^A$ is subsequently obtained through Eq.\eqref{eq13}. We present the equations used to calculate the theoretical epidemic thresholds at various levels in Table \ref{Table2}. 

In Fig.~\ref{fig5}(b) and (c), the theoretical epidemic thresholds at various levels are compared with simulation results obtained using the QS method under different exposed-to-infectious probability $\sigma$ and recovery probability $\mu$. It is observed that, except for the over-estimation by the macroscopic approach, all other methods underestimate the thresholds, with the mesoscopic method being the closest to the simulation results. The differences between the microscopic and microscopic approximations are minimal. Consequently, the mesoscopic results are adopted for subsequent theoretical analyses on the epidemic threshold.

\begin{table}
	\caption{$R_{0}^M$ at various levels}
	\label{Table2}
	\begin{ruledtabular}
		\begin{tabular}{c c} 
			Levels  & Formulas  \\ 
			\hline
			Microscopic &$\begin{aligned}
				R_0^M =& \Lambda_{\max}(H),\\
				h_{ij} =& \left(1-(1-\gamma_A)p_{i}^{A}\right)a_{ji}
			\end{aligned}$ \\
			Microscopic-approx & $R_0^M = \left(1-(1-\gamma_A)p^{A}\right)\Lambda_{\max}(A)$\\
			Mesoscopic & $R_0^M =\left(1-(1-\gamma_A)p^{A}\right)\frac{\langle k_1^2 \rangle}{\langle k_1 \rangle}$\\
			Macroscopic &$	R_0^M =\left(1-(1-\gamma_A)p^{A}\right) \langle k_1 \rangle$\\
			\hline
			$p^{A}_i$ & $\begin{aligned}
				p_{i}^{A} =& (1-r_{i})+(r_{i}-\delta) p_{i}^{A},\\
				r_{i}=&\left[1-k_g^m g_l\right]\prod_{j}\left[1-b_{ji}p_{j}^{A}\lambda\right]
			\end{aligned}$ \\
			$p^{A}$ &$p^{A} = \frac{1}{N} \sum_{i} p_{i}^{A}$\\
			
		\end{tabular}
	\end{ruledtabular}
\end{table}

The diffusion of the disease on the network is separately illustrated in Fig.~\ref{fig9} (a, b, c) for the case where the basic reproduction number $R_0 > 1$, and in Fig.~\ref{fig9} (d, e, f) for $R_0 < 1$. In Fig.~\ref{fig9}(a), the temporal diffusion of the disease is visualized, revealing that as time progresses, the disease spreads across various regions, with the central area being particularly prominent. When the diffusion stabilizes, the peripheral nodes remain healthy, as evidenced by the green nodes at Time = 10. Since peripheral nodes generally have smaller degrees, it can be inferred that nodes with lower degrees in the physical layer tend to remain healthier.  A more detailed discussion on the impact of network structure on the epidemic propagation will be provided in Sec.~\ref{networkstructure}. The emergence of awareness due to disease propagation and its subsequent diffusion is visually evident in Fig.~\ref{fig9}(b). Furthermore, the right panel of Fig.~\ref{fig9} (c) depicts the escalation of governmental attention from low to medium intensity in response to the epidemic's spread. The left and middle panels of Fig.~\ref{fig9}(c) compare the epidemic size and awareness size  over time using MMCA and MC simulations separately, demonstrating that MMCA aligns closely with MC, albeit slightly higher at steady state due to the lack of dynamical correlations among the states of neighbors in MMCA, which typically results in over-estimation\cite{granellDynamicalInterplayAwareness2013,Yu2025a}. For $R_0 < 1$, Fig.~\ref{fig9}(d) reveals a gradual decline in epidemic size leading to eventual extinction. Due to the initially large epidemic size, Fig.~\ref{fig9}(e) shows a rapid increase in awareness size, followed by stabilization. In the left and middle panels of Fig.~\ref{fig9}(f), the differences in epidemic and awareness sizes obtained using MMCA and MC methods are notably smaller compared to $R_0 > 1$. The right panel of Fig. 6(f) demonstrates the transition of governmental attention from high to medium and ultimately to low intensity as the epidemic diminishes and disappears. The results indicate that the MMCA method provides a robust approximation of epidemic diffusion, regardless of whether $R_0 > 1$ or $R_0 < 1$. Additionally, the attention of social nodes adapts dynamically in response to the spread of the epidemic.

\subsection{Influence of fundamental factors}
Influence of the infection probability $\beta$ and awareness probability $\lambda$ on the epidemic size and epidemic threshold is first investigated.
The exchange of disease information between individuals heightens awareness and prompts protective measures. Fig~\ref{fig4}(a) illustrates that, with a fixed $\lambda$, a higher $\beta$ correlates with a larger epidemic size $p^{EI}$, as indicated by the yellow arrow. The red line, representing a contour of the epidemic size, rises with increased $\lambda$, corresponding to a larger $\beta$. This suggests that if an epidemic is deemed safe when its scale falls below a certain threshold (zero being the epidemic threshold), a higher $\lambda$ can transform some unsafe diseases into safe ones, obviating the need for additional interventions. The more actively individuals exchange disease information, the greater $\lambda$ will be, thereby converting unsafe diseases into safe ones. Furthermore, Fig~\ref{fig4}(b) reveals that increases in $\beta$ and $\lambda$ all expand the scale of information, with $\lambda$'s effect being direct and $\beta$'s effect indirect, mediated by an increase in the number of infected individuals. A more detailed view is provided in the trajectory diagrams of  Fig.~\ref{fig4}(c) and  Fig.~\ref{fig4}(f).

In Fig.~\ref{fig5}(b) and (c), a higher exposed-to-infectious probability $\sigma$ (from state E to I) and recovery probability $\mu$ (from I to S) elevate the epidemic threshold, suggesting that shorter exposure periods ($1/\sigma$) and recovery periods ($1/\mu$) reduce epidemic outbreaks. This highlights the importance of proactive health measures, such as nucleic acid testing during pandemics, for early disease detection and intervention, promoting faster recovery and mitigating epidemic spread. These findings align with our theoretical result, i.e. $R_{0}^H = \frac{1}{\sigma}+\frac{1}{\mu_{H}}.$

\subsection{Influence of the social layer}

\begin{figure}
	\centering
	\includegraphics[width=0.8\linewidth]{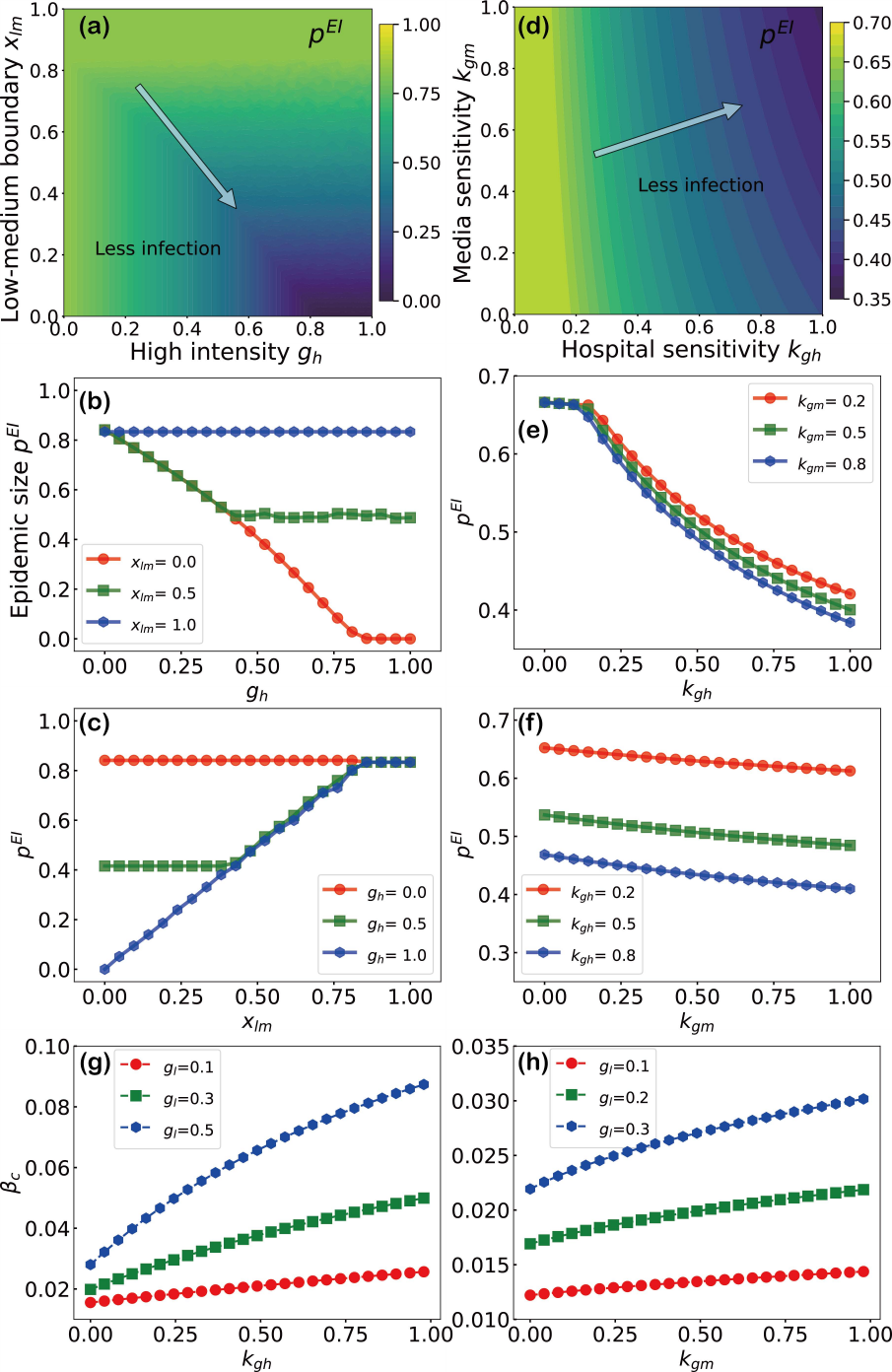}
	\caption{Impact of societal layers on the epidemic size and epidemic threshold. Panel (a) shows the phase diagram of the epidemic size with respect to the government feedback position $x_{lm}$ and feedback intensity $g_h$ (assuming $x_{lm} = x_{mh}$ and $g_l = 0$), while panels (b) and (c) depict trajectories of the epidemic size versus $x_{lm}$ and $g_h$. Panel (d) presents the phase diagram of the epidemic size with respect to the government influence on hospitals $k_{gh}$ and the media $k_{gm}$, with trajectories in panels (e) and (f). Panels (g) and (h) illustrate the effects of $k_{gh}$ and $k_{gm}$ on the epidemic threshold.}
	\label{fig7}
\end{figure}

To analyze the impact of the response intensity and sensitivity of the feedback function on the epidemic, we consider the situation in the absence of medium intensity (i.e., $x_{lm} = x_{mh}$). The response intensity, denoted by $g_h$, represents the strength of the government’s response to the epidemic, while the response sensitivity, represented by $x_{lm}$, reflects the government’s reaction based on the observed epidemic size. Fig~\ref{fig7}(a) illustrates the influence of $x_{lm}$ and $g_h$ on the epidemic size, revealing that a higher $g_h$ and a smaller $x_{lm}$ result in a smaller epidemic size. Fig~\ref{fig7}(b) shows that, for a given $x_{lm}$, increasing $g_h$ leads to a linear reduction in the epidemic size until it reaches $x_{lm}$. Conversely, Fig~\ref{fig7}(c) demonstrates that, for a fixed $g_h$, increasing $x_{lm}$ initially leaves the epidemic size unchanged, followed by a linear increase to the scale observed in the absence of government intervention. These findings indicate that a greater government sensitivity, characterized by earlier responses at lower $x_{lm}$, combined with higher response intensity, effectively reduces the epidemic size.

We further examine the influence of government on hospital and media nodes by analyzing the impact of parameters $k_{gh}$ and $k_{gm}$ on the epidemic size. Fig~\ref{fig7}(d) demonstrates that larger values of $k_{gh}$ and $k_{gm}$ correspond to a smaller epidemic size. Fig~\ref{fig7}(e) and (f) further reveal that $k_{gh}$ exerts a stronger influence compared to $k_{gm}$. These results indicate that the government-driven prioritization of disease by hospitals and media contributes to a better epidemic control, with hospitals playing a more significant role than the media. As depicted in Fig~\ref{fig7}(g) and (h), larger values of parameters $k_{gh}$ and $k_{gm}$ correspond to a higher epidemic threshold, and a larger $g_l$ also increases the epidemic threshold. This suggests that a greater focus the government puts on the prevention as well as a stronger influence on hospitals and the media, the more difficult an epidemic outbreak becomes possible.

\subsection{Influence of the network structure}
\label{networkstructure}
\begin{figure}
	\centering
	\includegraphics[width=1.0\linewidth]{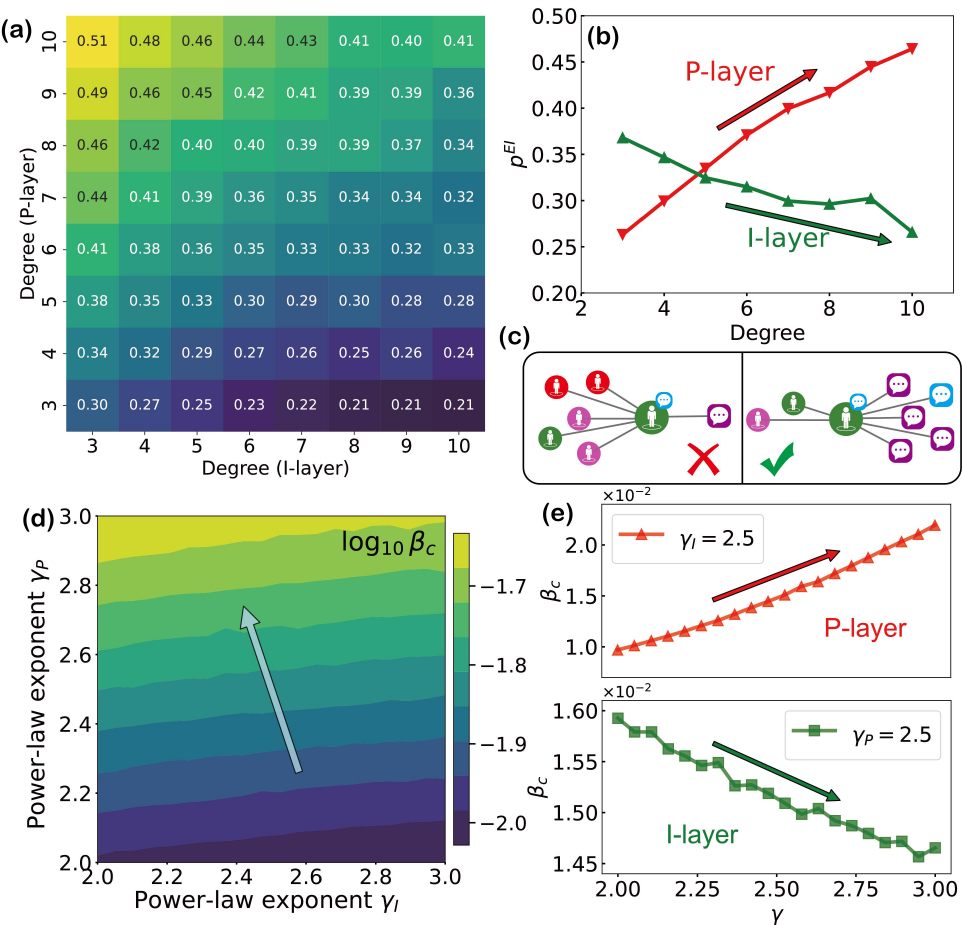}
	\caption{Impact of the network structure on the epidemic size and epidemic threshold. In a BA network, panels (a) and (b) depict the epidemic size across individuals with different degrees, considering both physical and information layers jointly (a) or separately (b). Panel (d) illustrates the epidemic threshold in UCM networks with varying exponents $\gamma_P$ and $\gamma_r$ for the physical and information layers, respectively. Panels (e) and (f) show the epidemic threshold as a function of $\gamma_P$ and $\gamma_r$. Panel (c) provides schematic representations of two population types: (left) information-isolated but frequently interacting, and (right) widely informed but non-interacting.}
	\label{fig8}
\end{figure}

\begin{figure}
	\centering
	\includegraphics[width=1.0\linewidth]{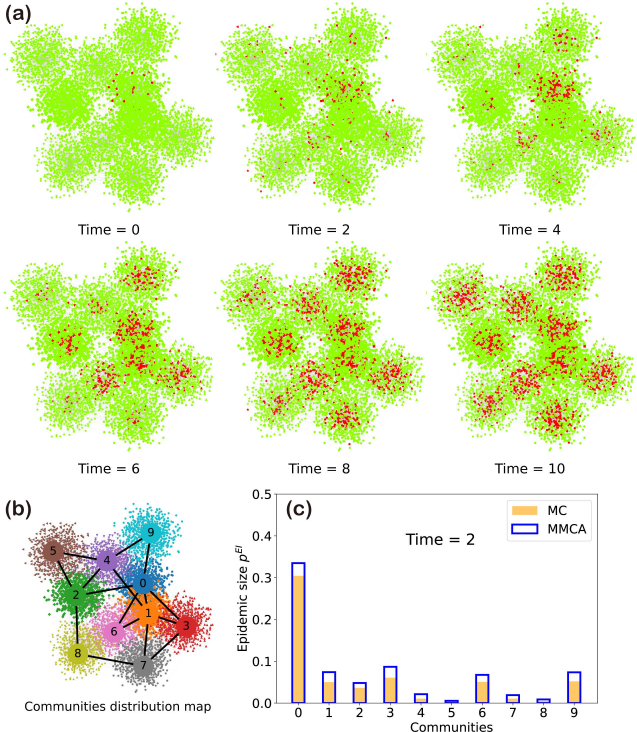}
	\caption{Epidemic spreading dynamics with community structure in the physical layer. (a)  shows the spreading of the epidemic across communities at times 0, 2, 4, 6, and 8.  (b) Inter-community connectivity relationships, generated using a BA network model. (c) Epidemic size within each community at time 2, calculated using the MMCA and MC  methods. The community network is constructed similarly to a BA network, with nodes connecting to neighboring communities with a probability of 0.1 and to the same community with a probability of 0.9.}
	\label{fig10}
\end{figure}

In this section, we examine the impact of population and community structures on the epidemic transmission. We model both physical and information layers using Barabási-Albert (BA) scale-free networks, which reflect real-world scenarios. For a given network structure, we analyze the epidemic size across different population groups. Fig.~\ref{fig8}(a) reveals that a higher degree in the information layer and a lower degree in the physical layer correspond to a lower infection rate. This is further detailed in Fig.~\ref{fig8}(b). These findings indicate that children, who frequently attend school, exhibit a higher degree in the physical layer but a lower degree in the information layer due to limited use of network devices, resulting in a higher infection rate. Thus, children are a critical group requiring continuously focused protection.

Furthermore, we investigate the influence of different community structures on the epidemic threshold using UCM networks with varying power-law exponents $\gamma_P$ and $\gamma_I$ for the physical and information layers, respectively, where $2 \leq \gamma_P, \gamma_I \leq 3$. A smaller exponent indicates a stronger network heterogeneity, with a higher likelihood of nodes with large degrees. Fig.~\ref{fig8}(d) demonstrates that a larger $\gamma_P$ and a smaller $\gamma_I$ lead to a higher epidemic threshold, as further elaborated in Fig.~\ref{fig8}(e). This suggests that the weaker heterogeneity in the physical layer and stronger heterogeneity in the information layer make epidemic outbreaks less likely. Compared to urban communities, rural communities exhibit a weaker heterogeneity in the physical layer, while the heterogeneity in the information layer is similar due to widespread internet access. Consequently, rural areas are relatively less prone to epidemic outbreaks.

Considering a more realistic network for the physical layer, which enjoys a community structure where nodes within the same community exhibit stronger connections, while connections between different communities are weaker. In Fig.~\ref{fig10}(b), we present the connectivity relationships among 10 communities and generate a network with such a community structure. Fig.~\ref{fig10}(a) illustrates the spread of the epidemic from Community 0 to other communities. Furthermore, Fig.~\ref{fig10}(c) shows the epidemic size within each community at time 2. It is observed that, apart from Community 0 where the epidemic first emerged, neighboring communities (1, 2, 3, 6, and 9) exhibit larger epidemic sizes compared to non-neighboring communities, indicating a fast diffusion to adjacent communities. This theoretical insight underscores the importance of implementing earlier and stronger containment measures in communities neighboring the initial outbreak.

\section{Conclusion}
In this study, we have developed a novel triplex network model for epidemic spreading that incorporates the influence of awareness layers and coupled social layers, including the government, the media, and hospital nodes. Our findings provide a theoretical foundation for understanding and controlling the epidemic spread in complex networks. Firstly, we derived the governing dynamic equations of epidemic transmission using the Microscopic Markov Chain Approach (MMCA), whose effectiveness is well validated through comparisons with Monte Carlo (MC) simulations. Using the uncorrelated configuration model (UCM) as an example, we obtained the theoretical epidemic thresholds at three levels: microscopic, mesoscopic, and macroscopic. Quasi-static (QS) simulations confirmed that the mesoscopic threshold most closely aligns with the simulated results. Secondly, we observed that the increased information dissemination probability in the awareness layer reduces the epidemic size, highlighting the importance of active information exchange among individuals in curbing epidemic spread. Thirdly, our results demonstrate that earlier and stronger government responses lead to reduced epidemic size. Similarly, increasing the government influence on the media and hospitals, particularly on the latter, contributes to an effective epidemic control. Fourthly, we found that with respect to fixed community structures, groups with more physical contacts and less information exchange, such as students in primary and secondary schools, exhibit higher infection rates. For different community structures, a weaker heterogeneity in the physical layer and a stronger heterogeneity in the information layer result in higher epidemic thresholds, making epidemic outbreaks less likely, as observed in rural areas with high internet penetration rates. We also simulate networks with community structures and observe that epidemics spread faster to adjacent communities. These findings offer valuable guidance for social epidemic prevention and control. Future research will focus on incorporating the cost of government feedback mechanisms and optimizing prevention and control strategies.

\begin{acknowledgments}
	This work was supported by the National Key R\&D Program of China (Grant No. 2023YFC2308702), Guangdong Basic and Applied Basic Research Foundation (2023A1515010157).
\end{acknowledgments}

\appendix
\section{Microscopic mean-field approach}
In this section, we conduct a theoretical analysis of the discrete microscopic mean-field equations introduced in the main text to derive the epidemic threshold. By introducing the time step $\Delta t$ and taking the limit as $\Delta t$ approaches zero, we obtain the continuous microscopic mean-field equations. Subsequently, we formally derive the basic reproduction number $R_0$ using the next-generation matrix method \cite{VanDenDriessche2002}, thereby establishing the epidemic threshold in the continuous case. Notably, the forms of the epidemic thresholds in both discrete and continuous settings are consistent.
\subsection{Epidemic threshold in discrete settings}
\label{proofs in MMCA}
When the system is in a stationary state, the probabilities for each state satisfy \(p_{i}(t+1) = p_{i}(t) = p_{i}\). Near the epidemic threshold, the probability of infection for each individual approaches zero, so we have \( p_{i}^{SU} \gg p_{i}^{EU} \), \( p_{i}^{SA} \gg p_{i}^{EA} + p_{i}^{IA} \), $g \approx g_l$ and $\lambda_{M} \approx k_g^m  g_l$. Additionally, since \( p_{i}^{U} = p_{i}^{SU} + p_{i}^{EU} \) and \( p_{i}^{A} = p_{i}^{SA} + p_{i}^{EA} + p_{i}^{IA} \), we find that 
\begin{equation}\label{approx1}
	p_{i}^{U} \approx p_{i}^{SU},\ p_{i}^{A} \approx p_{i}^{SA}.
\end{equation}
Assume that individuals in the state S first become individuals in the E upon infection, denoted as \(p_{i}^{E} = \epsilon_{i} \ll 1\). In the stationary state, according to Eq.~\eqref{MMCA}, we can obtain \[ p^{IA}_{i} = \frac{\sigma}{\mu_{H}^{0}} p^{E}_{i} = \frac{\sigma}{\mu_{H}^{0}} \epsilon_{i}.\] Therefore, we have the approximations 
\begin{equation*}
	\begin{aligned}
		q_{i}^{U} \approx& 1-\eta_{i}\beta_G,\\
		q_{i}^{A} \approx& 1-\eta_{i}\beta_{GA},
	\end{aligned}
\end{equation*}
where $\eta_{i}=(1+\frac{\sigma}{\mu_{H}^{0}} )\sum_{j}a_{ji}\epsilon_{j}$ and $\mu_{H}^{0} = \mu+k_g^h g_l-\mu k_g^h g_l$. Further referring Eq.~\eqref{approx1}, we have 
\begin{equation}\label{eq10}
	\begin{aligned}
		p_{i}^{U} &= r_{i}q_{i}^{U}p_{i}^{U}+\delta q_{i}^{U} p_{i}^{A}\approx r_{i}p_{i}^{U}+\delta p_{i}^{A},\\
		p_{i}^{A} &= (1-r_{i})q_{i}^{A}p_{i}^{U}+(1-\delta) q_{i}^{A} p_{i}^{A}\\
		&\approx(1-r_{i})p_{i}^{U}+(1-\delta) p_{i}^{A},\\
	\end{aligned}
\end{equation}
and 
\begin{equation*}
	\begin{aligned}
		\epsilon_{i} =& p_{i}^{EU}+p_{i}^{EA} \\
		=& (1-\sigma)\epsilon_{i}+\left(r_{i}p_{i}^{U} + \delta p_{i}^{A}\right)(1-q_{i}^{U}) \\
		&+ \left((1-r_{i})p_{i}^{U}+ (1-\delta)p_{i}^{A}\right)(1-q_{i}^{A}),\\
		=&(1-\sigma)\epsilon_{i}+\left(p_{i}^{U}\beta_G+p_{i}^{A}\beta_{GA}\right)\eta_{i}\\
		=&(1-\sigma)\epsilon_{i}+\left(1-(1-\gamma_A)p_{i}^{A}\right)\beta\eta_{i}(1 - g)\\
	\end{aligned}
\end{equation*}
Substituting \( \eta_{i} = \left(1 + \frac{\sigma}{\mu_{H}^{0}}\right) \sum_{j} a_{ji} \epsilon_{j} \), we can obtain
\begin{equation}\label{eq12}
	\begin{aligned}
		&\sigma\epsilon_{i} =\left(1-(1-\gamma_A)p_{i}^{A}\right)\beta(1 - g)(1+\frac{\sigma}{\mu_{H}^{0}} )\sum_{j}a_{ji}\epsilon_{j},\\
		&\sum_{j}\left\{\frac{\sigma\mu_{H}^{0}}{\beta(\sigma+\mu_{H}^{0})(1 - g_l)}\alpha_{ji} -\left(1-(1-\gamma_A)p_{i}^{A}\right)a_{ji}\right\}\epsilon_{j}=0,
	\end{aligned}
\end{equation}
where \(\alpha_{ji}\) is a delta function, defined as \(\alpha_{ji} = 1 \, (i = j)\) and \(\alpha_{ij} = 0 \, (i \neq j)\). Define the elements of matrix \( H \) as \( h_{ij} = \left(1-(1-\gamma_A)p_{i}^{A}\right)a_{ji} \), and let \( \epsilon = (\epsilon_1,\epsilon_2,\cdots)^T \). Then, Eq.~\eqref{eq12} can be expressed in the following matrix form:
\begin{equation}\label{eqH}
	H\epsilon = \frac{\sigma\mu_{H}^{0}}{\beta(\sigma+\mu_{H})(1 - g_l)}\epsilon.
\end{equation}
Thus, \( \frac{\sigma\mu_{H}^{0}}{\beta(\sigma+\mu_{H}^{0})} \) constitutes an eigenvalue of \( H \). Therefore, the epidemic threshold \( \beta_{c} \), defined as the minimum \( \beta \) that satisfies Eq.~\eqref{eqH}, reads 
\begin{equation*}
	\beta_{c} = \frac{\sigma\mu_{H}^{0}}{\Lambda_{\max}(H)(\sigma+\mu_{H}^{0})(1 - g_l)},
\end{equation*}
where \( \Lambda_{\max}(H)\) represents the largest eigenvalue of \( H \). According to the definition of \( H \), it depends on \( p_{i}^{A} \). From Eq.~\eqref{eq10} and $q_{i}^{A}+q_{i}^{U}=1$, we have 
\begin{equation}\label{Ap_eqpA}
	p_{i}^{A} = (1-r_{i})+(r_{i}-\delta) p_{i}^{A}.
\end{equation}

\begin{equation}\label{Ap_eqr}
	r_{i}=\underline{\left[1-k_g^m  g_l\right]}\prod_{j}\left[1-b_{ji}p_{j}^{A}\lambda\right].
\end{equation}
Therefore, we can iteratively obtain \( p_{i}^{A} \) based on Eq.~\eqref{Ap_eqpA} and Eq.~\eqref{Ap_eqr}, which allows us to derive \( H \), and subsequently determine \( \beta_{c} \).

\subsection{Continuous microscopic mean-field equations}
\label{CMMCA}
Based on Eq.~\eqref{MMCA}, we can derive the continuous microscopic mean-field equations. Introducing a time step of $\Delta t$ and rearranging Eq.~\eqref{MMCA}, we can take $\Delta t \to 0$ to obtain the following equations:

\begin{equation}\label{QME}
	\begin{aligned}
		\frac{d p_{i}^{SU}}{d t} =&-\sum_{j}(\lambda b_{ji}p_{j}^{A}p^{SU}_{i}+\beta_G a_{ji}p^{EI}_{j}p^{SU}_{i})-\lambda_M p^{SU}_{i} + \delta p_{i}^{SA},\\
		\frac{d p_{i}^{SA}}{d t} =& \sum_{j}(\lambda b_{ji}p_{j}^{A}p_{i}^{SU}- \beta_{GA} a_{ji}p^{EI}_{j}p^{SA}_{i}) +\lambda_M p^{SU}_{i}\\ & 
		- \delta p^{SA}_{i} +\mu_{H} p^{IA}_{i}, \\
		\frac{d p_{i}^{EU}}{d t} =& \sum_{j}(\beta_{G} a_{ji}p^{EI}_{j}p^{SU}_{i} - \lambda b_{ji}p_{j}^{A}p_{i}^{EU})
		-(\lambda_M + \sigma ) p^{EU}_{i}\\ & 
		+ \delta p^{EA}_{i}, \\
		\frac{d p_{i}^{EA}}{d t} =& \sum_{j}(\beta_{GA} a_{ji}p^{EI}_{j}p^{SA}_{i} + \lambda b_{ji}p_{j}^{A}p_{i}^{EU})
		+\lambda_M p^{EU}_{i}\\ & 
		- (\delta + \sigma) p^{EA}_{i}, \\
		\frac{d p_{i}^{IA}}{d t} =& \sigma (p_{i}^{EA}+ p_{i}^{EU}) -\mu_{H} p_{i}^{IA}.
	\end{aligned}
\end{equation}
From Eq.~\eqref{QME}, it is straightforward to verify the conservation law for the total probability:
$$
\frac{d}{dt} \left( p_{i}^{SU} + p_{i}^{SA} + p_{i}^{EU} + p_{i}^{EA} + p_{i}^{IA} \right) = 0.
$$

\subsection{Epidemic threshold in continuous settings}
\label{R0_QMF}

For the disease-free equilibrium (DFE), the basic reproduction number $R_0$ is calculated using the next-generation matrix method \cite{VanDenDriessche2002}. Let 
$$\mathcal{F} = \left(\begin{array}{c}
	\sum_{j}\beta_{G}a_{j1}p_{j}^{EI}p_{1}^{SU}\\
	\vdots\\
	\sum_{j}\beta_{G}a_{jN}p_{j}^{EI}p_{N}^{SU}\\
	\sum_{j}\beta_{GA}a_{j1}p_{j}^{EI}p_{1}^{SA}\\
	\vdots\\
	\sum_{j}\beta_{GA}a_{jN}p_{j}^{EI}p_{N}^{SA}\\
	0\\
	\vdots\\
	0
\end{array}\right)$$
denote the rate of new infections, and let
$$\mathcal{V} = \left(\begin{array}{c}
	\sum_{j}\lambda b_{j1}p_{j}^{A}p_{1}^{EU}+(\lambda_{M}+\sigma)p_{1}^{EU}-\delta p_{1}^{EA}\\
	\vdots\\
	\sum_{j}\lambda b_{jN}p_{j}^{A}p_{N}^{EU}+(\lambda_{M}+\sigma)p_{N}^{EU}-\delta p_{N}^{EA}\\
	-\sum_{j}\lambda b_{j1}p_{j}^{A}p_{1}^{EU}-\lambda_{M}p_{1}^{EU}+(\sigma+\delta) p_{1}^{EA}\\
	\vdots\\
	-\sum_{j}\lambda b_{jN}p_{j}^{A}p_{N}^{EU}-\lambda_{M}p_{N}^{EU}+(\sigma+\delta) p_{N}^{EA}\\
	-\sigma(p_{1}^{EU} + p_{1}^{EA})+\mu_{H}p_{1}^{IA}\\
	\vdots\\
	-\sigma(p_{N}^{EU} + p_{N}^{EA})+\mu_{H}p_{N}^{IA}\\
\end{array}\right)$$
represent the transitions between infected states.
Let $F$ and $V$ be the Jacobian matrices of $\mathcal{F}$ and $\mathcal{V}$, respectively, evaluated at the point $E^{0} = (p_{1}^{SU}, \dots, p_{N}^{SU}, p_{1}^{SA}, \dots, p_{N}^{SA}, 0, \dots, 0)$. Then, 
$$F = \beta(1-f_{obs}^g(0))\left(\begin{array}{ccc}
	A^TP^{U}& A^TP^{U} &A^TP^{U}\\
	\gamma_{A}A^TP^{A}& \gamma_{A}A^TP^{A}&\gamma_{A}A^TP^{A}\\
	\mathbf{0}&\mathbf{0}&\mathbf{0}
\end{array}\right),$$
where $P^{U} = \text{diag}(p_{1}^{U},\cdots,p_{N}^{U}), P^{A} = \mathbf{1}-P^{U}$,
and 
$$V = \left(\begin{array}{ccc}
	B+\sigma \mathbf{1} & -\delta \mathbf{1} &\mathbf{0}\\
	-B & (\delta + \sigma)\mathbf{1} &\mathbf{0}\\
	-\sigma \mathbf{1} & -\sigma \mathbf{1} &\mu_{H}^{0} \mathbf{1}\\
\end{array}\right),$$
where $B = \text{diag}(B_{1},\cdots,B_{N}),B_{i} = \sum_{j}b_{ji}p_{j}^{A}+k_g^m g_l,$ and $\mu_{H}^{0} = \mu+k_g^h g_l-\mu k_g^h g_l.$

Thus, we obtain 

\begin{widetext}
$$V^{-1} = \left(\begin{array}{ccc}
	\text{diag}(\frac{\delta+\sigma}{\sigma(B_1+\delta+\sigma)},\cdots,\frac{\delta+\sigma}{\sigma(B_N+\delta+\sigma)}) & \text{diag}(\frac{\delta}{\sigma(B_1+\delta+\sigma)},\cdots,\frac{\delta}{\sigma(B_N+\delta+\sigma)})  &\mathbf{0}\\
	\text{diag}(\frac{B_1}{\sigma(B_1+\delta+\sigma)},\cdots,\frac{B_N}{\sigma(B_N+\delta+\sigma)}) & \text{diag}(\frac{B_1+\sigma}{\sigma(B_1+\delta+\sigma)},\cdots,\frac{B_N+\sigma}{\sigma(B_N+\delta+\sigma)}) &\mathbf{0}\\
	1/\mu_{H}^{0} \mathbf{1} & 1/\mu_{H}^{0} \mathbf{1} &1/\mu_{H}^{0} \mathbf{1}\\
\end{array}\right),$$
\end{widetext}
and consequently, 
\begin{widetext}
$$FV^{-1}  =  \beta(1-f_{obs}^g(0)) \left(\begin{array}{ccc}
	(\frac{1}{\sigma}+\frac{1}{\mu_{H}^{0}})A^{T}P^{U}&(\frac{1}{\sigma}+\frac{1}{\mu_{H}^{0}})A^{T}P^{U}&\frac{1}{\mu_{H}^{0}}A^{T}P^{U}\\
	(\frac{1}{\sigma}+\frac{1}{\mu_{H}^{0}})\gamma_{A}A^{T}P^{A}&(\frac{1}{\sigma}+\frac{1}{\mu_{H}^{0}})\gamma_{A}A^{T}P^{A}&\frac{1}{\mu_{H}^{0}}\gamma_{A}A^{T}P^{A}\\
	\mathbf{0}&\mathbf{0}&\mathbf{0}
\end{array}\right).$$
\end{widetext}
Therefore, the dominant eigenvalue of $FV^{-1}$ coincides with that of 
$$ \beta(1-f_{obs}^g(0))(\frac{1}{\sigma}+\frac{1}{\mu_{H}^{0}})
\left(\begin{array}{cc}
	A^{T}P^{U}&A^{T}P^{U}\\
	\gamma_{A}A^{T}P^{A}&\gamma_{A}A^{T}P^{A}\\
\end{array}\right)
.$$

Hence, 
$$R_0 = \beta(1-f_{obs}^g(0))(\frac{1}{\sigma}+\frac{1}{\mu_{H}^{0}})\Lambda(H),$$
where $H = A^{T}(P^{U}+\gamma_{A}P^{A}).$ It is noteworthy that we have employed the same notation for the auxiliary matrix $H$ as used in MMCA, and it can be readily verified that they are identical. 

From the derived basic reproduction number $R_0$, we can obtain the epidemic threshold 
\begin{equation*}
	\beta_{c} = \frac{\sigma\mu_{H}^{0}}{\Lambda_{\max}(H)(\sigma+\mu_{H}^{0})(1 - g_l)},
\end{equation*} 
Remarkably, this result aligns precisely with the outcome derived from the MMCA, which is  a noteworthy observation.

\section{Mesoscopic mean-field approach}
\label{degreeMMCA}
Assuming that all nodes with identical degrees in both physical and information layers are statistically equivalent, we can employ the MMCA to derive degree-based mean-field equations, which we refer to as the discrete mesoscopic mean-field equations. Similar to the microscopic case, we further obtain the continuous mesoscopic mean-field equations and derive the epidemic thresholds for both discrete and continuous scenarios. Due to the similarity in methodology, the continuous case is omitted here. Notably, when the degree distributions of the physical and information layers are independent and both networks are uncorrelated configuration networks, the form of the epidemic threshold becomes significantly more concise.
\subsection{Discrete mesoscopic mean-field equations}
In this section, we consider a degree-based MMCA approach. First, we begin by assuming that each node (individual) is characterized by a degree vector $\mathbf{k} = (k_1, k_2)$, where $k_1$ and $k_2$ denote the node's degrees in the P-layer and I-layer, respectively. Here, $k_1$ and $k_2$ take discrete values $k_1 = 0, 1, 2, \dots, k_1^{\text{max}}$ and $k_2 = 0, 1, 2, \dots, k_2^{\text{max}}$, where $k_1^{\text{max}}$ and $k_2^{\text{max}}$ represent the maximum degrees in the P-layer and I-layer, respectively. Here, $P(\mathbf{k}) = P(k_1, k_2)$ denotes the degree distribution, i.e., the probability that a randomly selected node has degree $(k_1, k_2)$. Additionally, $P_1(\mathbf{k'} | \mathbf{k}) = P_1( k_1', k_2'|k_1, k_2 )$ and $P_2(\mathbf{k'} | \mathbf{k}) = P_2(k_1', k_2' | k_1, k_2)$ denote the conditional probabilities that, in the P-layer and I-layer respectively, a node with degree $\mathbf{k} = (k_1, k_2)$ connects via a randomly chosen edge to a node with degree $\mathbf{k} = (k_1', k_2')$. 

Assuming the topological connections of the P-layer and I-layer are independent, we have $P(\mathbf{k}) = P_{P}(k_1)P_{I}(k_2)$, where $P_{P}$ and $P_{I}$ are the degree distribution functions of the P-layer and I-layer, respectively. Further assuming that both layers are uncorrelated networks, the conditional probabilities are independent of the starting node, yielding 
$$P_1(\mathbf{k'}|\mathbf{k}) = \frac{k_1' P_{P}(k_1')P_{I}(k_2')}{\langle k_1' \rangle}$$ 
and 
$$P_2(\mathbf{k'}|\mathbf{k}) = \frac{k_2' P_{P}(k_1')P_{I}(k_2')}{\langle k_2' \rangle},$$
where $\langle k_1' \rangle$ and $\langle k_2' \rangle$ denote the average degrees of the P-layer and I-layer, respectively.

Let $p_{\mathbf{k}}^{SU}(t)$, $p_{\mathbf{k}}^{SA}(t)$, $p_{\mathbf{k}}^{EU}(t)$, $p_{\mathbf{k}}^{EA}(t)$, and $p_{\mathbf{k}}^{IA}(t)$ denote the probabilities that a node with degree $\mathbf{k}$ is in states SU, SA, EU, EA, and IA at time $t$, respectively. Additionally, $r_{\mathbf{k}}(t)$ represents the probability that a degree-$\mathbf{k}$ node in state U is not informed by any neighbors at time $t$, while $q_{\mathbf{k}}^{U}(t)$ and $q_{\mathbf{k}}^{A}(t)$ denote the probabilities that a degree-$\mathbf{k}$ node in SU or SA, respectively, is not infected by any neighbor at time $t$. From these definitions, we can readily derive the expressions:
\begin{equation}
	\begin{aligned}
		r_{\mathbf{k}}(t) =& (1-\lambda_{M})\left[1-\lambda\phi_2\right]^{k_2},\\
		q_{\mathbf{k}}^{U}(t) =& \left[1-\beta_{G}\phi_1\right]^{k_1},\\
		q_{\mathbf{k}}^{A}(t) =&\left[1-\beta_{GA}\phi_1\right]^{k_1},
	\end{aligned}
\end{equation}
where $$\phi_1 = \sum_{\mathbf{k^{\prime}}}P_1(\mathbf{k^{\prime}} | \mathbf{k}) p_{\mathbf{k^{\prime}}}^{EI}(t), \quad \phi_2(t) =\sum_{\mathbf{k^{\prime}}}P_2(\mathbf{k^{\prime}} | \mathbf{k}) p_{\mathbf{k^{\prime}}}^{A}(t),$$ $$p_{\mathbf{k^{\prime}}}^{A}(t) = p_{\mathbf{k^{\prime}}}^{SA}(t)+p_{\mathbf{k^{\prime}}}^{EA}(t)+p_{\mathbf{k^{\prime}}}^{IA}(t),$$
and 
$$p_{\mathbf{k^{\prime}}}^{EI}(t) = p_{\mathbf{k^{\prime}}}^{EU}(t)+p_{\mathbf{k^{\prime}}}^{EA}(t)+p_{\mathbf{k^{\prime}}}^{IA}(t).$$

The observation function $obs(t)$ is defined as:
$$obs(t) = \sum_{\mathbf{k}} P(\mathbf{k})p^{EI}_{\mathbf{k}}(t).$$

Based on the Markov state transition tree, we can derive the corresponding discrete mesoscopic mean-field equations: 
\begin{equation}\label{degree-based MMCA}
	\begin{aligned}
		p_{\mathbf{k}}^{SU}(t+1) =& r_{\mathbf{k}}(t)q_{\mathbf{k}}^{U}(t)p_{\mathbf{k}}^{SU}(t)+\delta q_{\mathbf{k}}^{U}(t)p_{\mathbf{k}}^{SA}(t) \\&+ \delta \mu_{H} p_{\mathbf{k}}^{IA}(t),\\
		p_{\mathbf{k}}^{SA}(t+1) =& (1-r_{\mathbf{k}}(t))q_{\mathbf{k}}^{A}(t)p_{\mathbf{k}}^{SU}(t)\\&+(1-\delta) q_{\mathbf{k}}^{A}(t)p_{\mathbf{k}}^{SA}(t)\\& + (1-\delta) \mu_{H} p_{\mathbf{k}}^{IA}(t), \\
		p_{\mathbf{k}}^{EU}(t+1) =& r_{\mathbf{k}}(t)(1-q_{\mathbf{k}}^{U}(t))p_{\mathbf{k}}^{SU}(t)\\&+\delta (1-q_{\mathbf{k}}^{U}(t))p_{\mathbf{k}}^{SA}(t)\\& + r_{\mathbf{k}}(t)(1-\sigma)p_{\mathbf{k}}^{EU}(t) \\&+ \delta(1-\sigma)p_{\mathbf{k}}^{EA}(t),\\
		p_{\mathbf{k}}^{EA}(t+1) =& (1-r_{\mathbf{k}}(t))(1-q_{\mathbf{k}}^{A}(t))p_{\mathbf{k}}^{SU}(t)\\&+(1-\delta) (1-q_{\mathbf{k}}^{A}(t))p_{\mathbf{k}}^{SA}(t) \\
		&+ (1-r_{\mathbf{k}}(t))(1-\sigma)p_{\mathbf{k}}^{EU}(t) \\&+ (1-\delta)(1-\sigma)p_{\mathbf{k}}^{EA}(t),\\
		p_{\mathbf{k}}^{IA}(t+1) =& \sigma p_{\mathbf{k}}^{EA}(t)+\sigma p_{\mathbf{k}}^{EU}(t)+(1-\mu_{H}) p_{\mathbf{k}}^{IA}(t).
	\end{aligned}
\end{equation}

\subsection{Continuous mesoscopic mean-field equations}
Based on Eq.~\eqref{degree-based MMCA}, we can derive the continuous mesoscopic mean-field equations, also termed the heterogeneous mean-field equations. Considering a time step of $\Delta t$ and rearranging Eq.~\eqref{degree-based MMCA}, we can take $\Delta t \to 0$ to obtain the following equations:
\begin{equation}\label{HME}
	\begin{aligned}
		\frac{d p_{\mathbf{k}}^{SU}}{d t} =&-k_{2}\lambda \phi_2 p^{SU}_{\mathbf{k}}-k_{1} \beta_G \phi_1 p^{SU}_{\mathbf{k}}-\lambda_M p^{SU}_{\mathbf{k}} + \delta p_{\mathbf{k}}^{SA},\\
		\frac{d p_{\mathbf{k}}^{SA}}{d t} =& k_{2}\lambda \phi_2 p_{\mathbf{k}}^{SU} -k_{1} \beta_{GA} \phi_1 p^{SA}_{\mathbf{k}} +\lambda_M p^{SU}_{\mathbf{k}} \\ &- \delta p^{SA}_{\mathbf{k}} +\mu_{H} p^{IA}_{\mathbf{k}}, \\
		\frac{d p_{\mathbf{k}}^{EU}}{d t} =& k_{1} \beta_G \phi_1 p^{SU}_{\mathbf{k}} -k_{2}\lambda \phi_2 p_{\mathbf{k}}^{EU}
		-(\lambda_M + \sigma ) p^{EU}_{\mathbf{k}}\\ &+ \delta p^{EA}_{\mathbf{k}}, \\
		\frac{d p_{\mathbf{k}}^{EA}}{d t} =& k_{1} \beta_{GA} \phi_1 p^{SA}_{\mathbf{k}} + k_{2}\lambda \phi_2p_{\mathbf{k}}^{EU}+\lambda_M p^{EU}_{\mathbf{k}}\\ & - (\delta + \sigma) p^{EA}_{\mathbf{k}}, \\
		\frac{d p_{\mathbf{k}}^{IA}}{d t} =& \sigma (p_{\mathbf{k}}^{EA}+ p_{\mathbf{k}}^{EU}) -\mu_{H} p_{\mathbf{k}}^{IA}.
	\end{aligned}
\end{equation}
From Eq.~\eqref{HME}, it is straightforward to verify the conservation law for the total probability:
$$
\frac{d}{dt} \left( p_{\mathbf{k}}^{SU} + p_{\mathbf{k}}^{SA} + p_{\mathbf{k}}^{EU} + p_{\mathbf{k}}^{EA} + p_{\mathbf{k}}^{IA} \right) = 0.
$$

\subsection{Epidemic threshold in discrete settings}
\label{degreebasedMMCA}
When the system is in a stationary state, the probabilities for each state satisfy \(p_{\mathbf{k}}(t+1) = p_{\mathbf{k}}(t) = p_{\mathbf{k}}\). Near the epidemic threshold, the probability of infection for each individual approaches zero, so we have \( p_{\mathbf{k}}^{SU} \gg p_{\mathbf{k}}^{EU} \), \( p_{\mathbf{k}}^{SA} \gg p_{\mathbf{k}}^{EA} + p_{\mathbf{k}}^{IA} \), $g \approx g_l$ and $\lambda_{M} \approx k_g^m  g_l$. Additionally, since \( p_{\mathbf{k}}^{U} = p_{\mathbf{k}}^{SU} + p_{\mathbf{k}}^{EU} \) and \( p_{\mathbf{k}}^{A} = p_{\mathbf{k}}^{SA} + p_{\mathbf{k}}^{EA} + p_{\mathbf{k}}^{IA} \), we find that 
\begin{equation}\label{approx1 degree-based}
	p_{\mathbf{k}}^{U} \approx p_{\mathbf{k}}^{SU},\ p_{\mathbf{k}}^{A} \approx p_{\mathbf{k}}^{SA}.
\end{equation}
Assume that individuals in the state S first become individuals in the E upon infection, denoted as \(p_{\mathbf{k}}^{E} = \epsilon_{\mathbf{k}} \ll 1\). In the stationary state, according to Eq.~\eqref{degree-based MMCA}, we can obtain \[ p^{IA}_{\mathbf{k}} = \frac{\sigma}{\mu_{H}^{0}} p^{E}_{\mathbf{k}} = \frac{\sigma}{\mu_{H}^{0}} \epsilon_{\mathbf{k}}.\] Therefore, we have the approximations 
\begin{equation*}
	\begin{aligned}
		q_{\mathbf{k}}^{U} \approx& 1-\eta_{\mathbf{k}}\beta_G,\\
		q_{\mathbf{k}}^{A} \approx& 1-\eta_{\mathbf{k}}\beta_{GA},
	\end{aligned}
\end{equation*}
where $\eta_{\mathbf{k}}=k_1(1+\frac{\sigma}{\mu_{H}^{0}} )\sum_{\mathbf{k^{\prime}}}P_1(\mathbf{k^{\prime}}|\mathbf{k})\epsilon_{\mathbf{k^{\prime}}}$ and $\mu_{H}^{0} = \mu+k_g^h g_l-\mu k_g^h g_l$. In the stationary state, according to Eq.~\eqref{degree-based MMCA} and Eq.~\eqref{approx1 degree-based}, we can obtain 
\begin{equation}\label{eq10 degree-based}
	\begin{aligned}
		p_{\mathbf{k}}^{U} &= r_{\mathbf{k}}q_{\mathbf{k}}^{U}p_{\mathbf{k}}^{U}+\delta q_{\mathbf{k}}^{U} p_{\mathbf{k}}^{A}\approx r_{\mathbf{k}}p_{\mathbf{k}}^{U}+\delta p_{\mathbf{k}}^{A},\\
		p_{\mathbf{k}}^{A} &= (1-r_{\mathbf{k}})q_{\mathbf{k}}^{A}p_{\mathbf{k}}^{U}+(1-\delta) q_{\mathbf{k}}^{A} p_{\mathbf{k}}^{A}\\
		&\approx(1-r_{\mathbf{k}})p_{\mathbf{k}}^{U}+(1-\delta) p_{\mathbf{k}}^{A},\\
	\end{aligned}
\end{equation}
and 
\begin{equation*}
	\begin{aligned}
		\epsilon_{\mathbf{k}} =& p_{\mathbf{k}}^{EU}+p_{\mathbf{k}}^{EA} \\
		=& (1-\sigma)\epsilon_{\mathbf{k}}+\left(r_{\mathbf{k}}p_{\mathbf{k}}^{U} + \delta p_{\mathbf{k}}^{A}\right)(1-q_{\mathbf{k}}^{U}) \\
		&+ \left((1-r_{\mathbf{k}})p_{\mathbf{k}}^{U}+ (1-\delta)p_{\mathbf{k}}^{A}\right)(1-q_{\mathbf{k}}^{A}),\\
		=&(1-\sigma)\epsilon_{\mathbf{k}}+\left(p_{\mathbf{k}}^{U}\beta_G+p_{\mathbf{k}}^{A}\beta_{GA}\right)\eta_{\mathbf{k}}\\
		=&(1-\sigma)\epsilon_{\mathbf{k}}+\left(1-(1-\gamma_A)p_{\mathbf{k}}^{A}\right)\beta\eta_{\mathbf{k}}(1 - g)\\
	\end{aligned}
\end{equation*}
Substituting \( \eta_{\mathbf{k}}=k_1(1+\frac{\sigma}{\mu_{H}^{0}} )\sum_{\mathbf{k^{\prime}}}P_1(\mathbf{k^{\prime}}|\mathbf{k})\epsilon_{\mathbf{k^{\prime}}} \), we can obtain
\begin{widetext}
	$$
\begin{aligned}
	&\sigma\epsilon_{\mathbf{k}} =\left(1-(1-\gamma_A)p_{\mathbf{k}}^{A}\right)\beta(1 - g)k_1(1+\frac{\sigma}{\mu_{H}^{0}} )\sum_{\mathbf{k^{\prime}}}P_1(\mathbf{k^{\prime}}|\mathbf{k})\epsilon_{\mathbf{k^{\prime}}} ,\label{eq9 degree-based}\\
	&\sum_{\mathbf{k^{\prime}}}\left\{\frac{\sigma\mu_{H}^{0}}{\beta(\sigma+\mu_{H})(1 - g_l)}\alpha_{\mathbf{k^{\prime}}\mathbf{k}} -\left(1-(1-\gamma_A)p_{\mathbf{k}}^{A}\right)k_1P_1(\mathbf{k^{\prime}}|\mathbf{k})\right\}\epsilon_{\mathbf{k^{\prime}}}=0,\label{eq12 degree-based}
\end{aligned}$$
\end{widetext}
where \(\alpha_{ji}\) is a delta function, defined as \(\alpha_{\mathbf{k^{\prime}}\mathbf{k}} = 1 \, (\mathbf{k^{\prime}}=\mathbf{k})\) and \(\alpha_{\mathbf{k^{\prime}}\mathbf{k}} = 0 \, (\mathbf{k^{\prime}}\neq \mathbf{k})\). Define the elements of matrix \( H \) as \( h_{\mathbf{k^{\prime}}\mathbf{k}} = \left(1-(1-\gamma_A)p_{\mathbf{k}}^{A}\right)k_1P_1(\mathbf{k^{\prime}}|\mathbf{k}) \), and let \( \epsilon = (\epsilon_{\mathbf{k_1}},\epsilon_{\mathbf{k_2}},\cdots)^T \). Then, Eq.~\eqref{eq12 degree-based} can be expressed in the following matrix form:
\begin{equation}\label{eqH degree-based}
	H\epsilon = \frac{\sigma\mu_{H}^{0}}{\beta(\sigma+\mu_{H}^{0})(1 - g_l)}\epsilon.
\end{equation}
Thus, \( \frac{\sigma\mu_{H}^{0}}{\beta(\sigma+\mu_{H}^{0})} \) constitutes an eigenvalue of \( H \). Therefore, the epidemic threshold \( \beta_{c} \), defined as the minimum \( \beta \) that satisfies Eq.~\eqref{eqH degree-based}, is given by 
\begin{equation*}
	\beta_{c} = \frac{\sigma\mu_{H}^{0}}{\Lambda_{\max}(H)(\sigma+\mu_{H}^{0})(1 - g_l)},
\end{equation*}
or
\begin{equation*}
	\beta_{c} = \frac{\sigma\mu_{H}^{0}}{\Lambda_{\max}(H)(\sigma+\mu_{H}^{0})(1 - g_l)},
\end{equation*}
where \( \Lambda_{\max}(H)\) represents the largest eigenvalue of \( H \). According to the definition of \( H \), it depends on \( p_{i}^{A} \). From Eq.~\eqref{eq10 degree-based} and $q_{\mathbf{k}}^{A}+q_{\mathbf{k}}^{U}=1$, we have 
\begin{equation}\label{Ap_eqpA degree-based}
	p_{\mathbf{k}}^{A} = (1-r_{\mathbf{k}})+(r_{\mathbf{k}}-\delta) p_{\mathbf{k}}^{A}.
\end{equation}

\begin{equation}\label{Ap_eqr degree-based}
	r_{\mathbf{k}}=\underline{\left[1-k_g^m  g_l\right]}\left[1-\lambda \sum_{\mathbf{k^{\prime}}}P_2(\mathbf{k^{\prime}} | \mathbf{k}) p_{\mathbf{k^{\prime}}}^{A}(t)\right]^{k_2}.
\end{equation}

Let $\Phi = \sum_{\mathbf{k^{\prime}}} P_2(\mathbf{k^{\prime}} | \mathbf{k}) p_{\mathbf{k^{\prime}}}^{A}(t)$, it follows that:
\begin{equation}\label{PA degree-based}
	p_{\mathbf{k}}^A = \frac{1 - \left[1 - k_g^m  g_l\right] \left[1 - \lambda \Phi\right]^{k_2}}{\delta + 1 - \left[1 - k_g^m  g_l\right] \left[1 - \lambda \Phi\right]^{k_2}},
\end{equation}
and consequently:
\begin{equation}\label{Phi degree-based}
	\Phi = \sum_{\mathbf{k^{\prime}}} P_2(\mathbf{k^{\prime}} | \mathbf{k}) \frac{1 - \left[1 - k_g^m  g_l\right] \left[1 - \lambda \Phi\right]^{k_2^{\prime}}}{\delta + 1 - \left[1 - k_g^m  g_l\right] \left[1 - \lambda \Phi\right]^{k_2^{\prime}}}.
\end{equation}
Therefore, we can iteratively compute $\Phi$ based on Eq.\eqref{Phi degree-based}, and subsequently substitute it into Eq.\eqref{PA degree-based} to obtain $p_{\mathbf{k}}^{A}$. This enables the derivation of $H$, and ultimately facilitates the determination of the critical threshold $\beta_{c}$.  Note that the term $p_{\mathbf{k}}^A$ here depends solely on the degree of the I-layer.

\subsection{Epidemic threshold in the uncorrelated network} 
Assuming the topological connections of the P-layer and I-layer are independent, we have $P(\mathbf{k}) = P_{P}(k_1)P_{I}(k_2)$, where $P_{P}$ and $P_{I}$ are the degree distribution functions of the P-layer and I-layer, respectively. Further assuming that both layers are uncorrelated networks, the conditional probabilities are independent of the starting node, yielding 
$$P_1(\mathbf{k'}|\mathbf{k}) = \frac{k_1' P_{P}(k_1')P_{I}(k_2')}{\langle k_1' \rangle}$$ 
and 
$$P_2(\mathbf{k'}|\mathbf{k}) = \frac{k_2' P_{P}(k_1')P_{I}(k_2')}{\langle k_2' \rangle},$$
where $\langle k_1' \rangle$ and $\langle k_2' \rangle$ denote the average degrees of the P-layer and I-layer, respectively.
From \eqref{eq9 degree-based}, and under the assumption of an uncorrelated network, it follows that 
$$
\begin{aligned}
	&\sigma\sum_{k_1}\sum_{k_2}k_1P_{P}(k_1)P_{I}(k_2)\epsilon_{\mathbf{k}}\\ &=\left(\langle k_1^2 \rangle-(1-\gamma_A)\sum_{k_1}\sum_{k_2}{k_1}^2P_{P}(k_1)P_{I}(k_2)p_{\mathbf{k}}^{A}\right)\times\\&\beta(1 - g) (1+\frac{\sigma}{\mu_{H}^{0}} )\sum_{k_{1}^{'}}\sum_{k_{2}^{'}} \frac{k_1^{\prime} P_{P}(k_1^{\prime})P_{I}(k_2^{\prime})}{\langle k_1^{\prime} \rangle} \epsilon_{\mathbf{k^{\prime}}}\\
	&\sigma =\left(\langle k_1^2 \rangle-(1-\gamma_A)\sum_{k_1}\sum_{k_2}{k_1}^2P_{P}(k_1)P_{I}(k_2)p_{\mathbf{k}}^{A}\right)\times\\&\beta(1 - g) (1+\frac{\sigma}{\mu_{H}^{0}} ) \frac{1}{\langle k_1 \rangle}\\
\end{aligned}
.$$
From Eq.~\eqref{PA degree-based} , it can be observed that $p_{\mathbf{k}}^{A}$ is independent of the degree $k_1$ in the P-layer, and we denote $p_{\mathbf{k}}^A = p_{k_2}^{A}$. This implies that $$\sum_{k_2} P_{I}(k_2) p_{\mathbf{k}}^{A} = \sum_{k_1} \sum_{k_2} P_{P}(k_1) P_{I}(k_2) p_{\mathbf{k}}^{A},\quad \forall k_1$$ which can be denoted as $p^A$.  Consequently, the aforementioned expression can be further simplified to yield 
$$
\begin{aligned}
	&\sigma = \left(1-(1-\gamma_A)p^{A}\right)\beta(1 - g) (1+\frac{\sigma}{\mu_{H}^{0}} ) \frac{\langle k_1^2 \rangle}{\langle k_1 \rangle}\\
\end{aligned}.
$$
Therefore, the epidemic threshold is 
\begin{equation*}
	\beta_{c} = \frac{\sigma\mu_{H}^{0}}{(\sigma+\mu_{H}^{0})(1 - g_l)}\frac{\langle k_1 \rangle}{\langle k_1^2 \rangle}\frac{1}{\left(1-(1-\gamma_A)p^{A}\right)}.
\end{equation*}

Based on equations \eqref{PA degree-based} and \eqref{Phi degree-based}, we can derive the expression 
\begin{equation*}
	\Phi =\frac{1}{\langle k_2 \rangle} \sum_{k_2}k_2 P_{I}(k_2)\frac{1 - \left[1 - k_g^m  g_l\right] \left[1 - \lambda \Phi\right]^{k_2}}{\delta + 1 - \left[1 - k_g^m  g_l\right] \left[1 - \lambda \Phi\right]^{k_2}}.
\end{equation*}
Subsequently, by iteratively solving for $\Phi$, we substitute $\Phi$ into the equation
\begin{equation*}
	p_{k_2}^A = \frac{1 - \left[1 - k_g^m  g_l\right] \left[1 - \lambda \Phi\right]^{k_2}}{\delta + 1 - \left[1 - k_g^m  g_l\right] \left[1 - \lambda \Phi\right]^{k_2}},
\end{equation*}
to obtain $p_{k_2}^{A}$. Finally, the overall probability $p^{A}$ is determined by summing over all degree classes $k_2$ as $$p^{A} = \sum_{k_2} P_{I}(k_2) p_{k_2}^{A}.$$

Another approach to obtain the epidemic threshold for an uncorrelated network is to directly compute $\Lambda_{\max}(H)$. 
In this setting, the elements of $H$ are given by 
$$h_{\mathbf{k}\mathbf{k'}} = \left(1-(1-\gamma_A)p_{\mathbf{k}}^{A}\right)k_1\frac{k_1' P_{P}(k_1')P_{I}(k_2')}{\langle k_1' \rangle}.$$
Thus, $H = \mathbf{a} \otimes \mathbf{b}^T,$
where $\mathbf{a} =\left\{ \left(1-(1-\gamma_A)p_{\mathbf{k}}^{A}\right)k_1 \right\}_{\mathbf{k}}$ and $\mathbf{b} =\left\{ \frac{k_1 P_{P}(k_1)P_{I}(k_2)}{\langle k_1 \rangle}\right\}_{\mathbf{k}}$. Consequently, the maximum eigenvalue of $H$, denoted as $\Lambda_{\max}(H)$, is equal to the inner product of $\mathbf{a}$ and $\mathbf{b}$, that is, 
\begin{widetext}
$$
\begin{aligned}
	\Lambda_{\max}(H) =& \mathbf{a}^T \mathbf{b} \\
	=& \sum_{k_1}\sum_{k_2}\left(1-(1-\gamma_A)p_{\mathbf{k}}^{A}\right)k_1\frac{k_1 P_{P}(k_1)P_{I}(k_2)}{\langle k_1 \rangle} \\
	=&\frac{1}{\langle k_1 \rangle} \sum_{k_1}\left(1-(1-\gamma_A)\sum_{k_2}p_{\mathbf{k}}^{A}P_{I}(k_2)\right)k_1^2P_{P}(k_1)\\
	=&\frac{1}{\langle k_1 \rangle} \sum_{k_1}\left(1-(1-\gamma_A)\sum_{k_2}p_{k_2}^{A}P_{I}(k_2)\right)k_1^2P_{P}(k_1)\\
	=&\frac{\langle k_1^2 \rangle}{\langle k_1 \rangle} \left(1-(1-\gamma_A)p^{A}\right)
	.
\end{aligned}
$$
\end{widetext}
In the above procedure, $p_{\mathbf{k}}^{A}$ is replaced by $p_{k_2}^{A}$ due to Eq.~\eqref{PA degree-based}.

\section{Macroscopic mean-field approach}
\label{MacroscopicMMCA}
Assuming all nodes are statistically equivalent, thereby neglecting the network heterogeneity, we can set $a_{ij} = \frac{2\langle k_1 \rangle}{N-1}$ and $b_{ij} = \frac{2\langle k_2 \rangle}{N-1}$ in the microscopic mean-field equations. Let $p^{SU} = \frac{1}{N}\sum_{i} p_{i}^{SU}$, with analogous definitions for $p^{SA}$, $p^{EU}$, $p^{EA}$, and $p^{IA}$. Under these conditions and based on Eq.~\eqref{QME}, we can derive the mean-field equations as follows:
\begin{equation}\label{ME}
	\begin{aligned}
		\frac{d p^{SU}}{d t} =&-2\lambda \langle k_{2} \rangle p^{SU}p^{A}- 2\beta_G \langle k_{1} \rangle p^{SU}p^{EI}-\lambda_M p^{SU} + \delta p^{SA},\\
		\frac{d p^{SA}}{d t} =& 2\lambda \langle k_{2} \rangle p^{SU}p^{A} - 2 \beta_{GA} \langle k_{1} \rangle p^{SA}p^{EI} +\lambda_M p^{SU} - \delta p^{SA} \\ &+\mu_{H} p^{IA}, \\
		\frac{d p^{EU}}{d t} =&  - 2\lambda \langle k_{2} \rangle p^{EU}p^{A} + 2\beta_G \langle k_{1} \rangle p^{SU}p^{EI} 
		-(\lambda_M + \sigma ) p^{EU}\\ &+ \delta p^{EA}, \\
		\frac{d p^{EA}}{d t} =&  2\lambda \langle k_{2} \rangle p^{EU}p^{A} +  2 \beta_{GA} \langle k_{1} \rangle p^{SA}p^{EI}+\lambda_M p^{EU}\\ & - (\delta + \sigma) p^{EA}, \\
		\frac{d p^{IA}}{d t} =& \sigma (p^{EA}+ p^{EU}) -\mu_{H} p^{IA},
	\end{aligned}
\end{equation}
where  $p^{A}= p^{SA}+p^{EA}+p^{IA}$
and 
$p^{EI} = p^{EU}+p^{EA}+p^{IA}.$
\nocite{*}

\bibliography{reference}

\begin{thebibliography}{37}%
\makeatletter
\providecommand \@ifxundefined [1]{%
 \@ifx{#1\undefined}
}%
\providecommand \@ifnum [1]{%
 \ifnum #1\expandafter \@firstoftwo
 \else \expandafter \@secondoftwo
 \fi
}%
\providecommand \@ifx [1]{%
 \ifx #1\expandafter \@firstoftwo
 \else \expandafter \@secondoftwo
 \fi
}%
\providecommand \natexlab [1]{#1}%
\providecommand \enquote  [1]{``#1''}%
\providecommand \bibnamefont  [1]{#1}%
\providecommand \bibfnamefont [1]{#1}%
\providecommand \citenamefont [1]{#1}%
\providecommand \href@noop [0]{\@secondoftwo}%
\providecommand \href [0]{\begingroup \@sanitize@url \@href}%
\providecommand \@href[1]{\@@startlink{#1}\@@href}%
\providecommand \@@href[1]{\endgroup#1\@@endlink}%
\providecommand \@sanitize@url [0]{\catcode `\\12\catcode `\$12\catcode
  `\&12\catcode `\#12\catcode `\^12\catcode `\_12\catcode `\%12\relax}%
\providecommand \@@startlink[1]{}%
\providecommand \@@endlink[0]{}%
\providecommand \url  [0]{\begingroup\@sanitize@url \@url }%
\providecommand \@url [1]{\endgroup\@href {#1}{\urlprefix }}%
\providecommand \urlprefix  [0]{URL }%
\providecommand \Eprint [0]{\href }%
\providecommand \doibase [0]{https://doi.org/}%
\providecommand \selectlanguage [0]{\@gobble}%
\providecommand \bibinfo  [0]{\@secondoftwo}%
\providecommand \bibfield  [0]{\@secondoftwo}%
\providecommand \translation [1]{[#1]}%
\providecommand \BibitemOpen [0]{}%
\providecommand \bibitemStop [0]{}%
\providecommand \bibitemNoStop [0]{.\EOS\space}%
\providecommand \EOS [0]{\spacefactor3000\relax}%
\providecommand \BibitemShut  [1]{\csname bibitem#1\endcsname}%
\let\auto@bib@innerbib\@empty
\bibitem [{\citenamefont {Malik}\ \emph {et~al.}(2024)\citenamefont {Malik},
  \citenamefont {Asghar},\ and\ \citenamefont
  {Waheed}}]{malikOutliningRecentUpdates2024}%
  \BibitemOpen
  \bibfield  {author} {\bibinfo {author} {\bibfnamefont {S.}~\bibnamefont
  {Malik}}, \bibinfo {author} {\bibfnamefont {M.}~\bibnamefont {Asghar}},\ and\
  \bibinfo {author} {\bibfnamefont {Y.}~\bibnamefont {Waheed}},\ }\bibfield
  {title} {\bibinfo {title} {Outlining recent updates on influenza therapeutics
  and vaccines: {{A}} comprehensive review},\ }\href
  {https://doi.org/10.1016/j.jvacx.2024.100452} {\bibfield  {journal} {\bibinfo
   {journal} {Vaccine: X}\ }\textbf {\bibinfo {volume} {17}},\ \bibinfo {pages}
  {100452} (\bibinfo {year} {2024})}\BibitemShut {NoStop}%
\bibitem [{\citenamefont {Platto}\ \emph {et~al.}(2020)\citenamefont {Platto},
  \citenamefont {Xue},\ and\ \citenamefont
  {Carafoli}}]{plattoCOVID19AnnouncedPandemic2020}%
  \BibitemOpen
  \bibfield  {author} {\bibinfo {author} {\bibfnamefont {S.}~\bibnamefont
  {Platto}}, \bibinfo {author} {\bibfnamefont {T.}~\bibnamefont {Xue}},\ and\
  \bibinfo {author} {\bibfnamefont {E.}~\bibnamefont {Carafoli}},\ }\bibfield
  {title} {\bibinfo {title} {{{COVID19}}: An announced pandemic},\ }\href
  {https://doi.org/10.1038/s41419-020-02995-9} {\bibfield  {journal} {\bibinfo
  {journal} {Cell Death \& Disease}\ }\textbf {\bibinfo {volume} {11}},\
  \bibinfo {pages} {799} (\bibinfo {year} {2020})}\BibitemShut {NoStop}%
\bibitem [{\citenamefont {Dietz}\ and\ \citenamefont
  {Heesterbeek}(2002)}]{dietzDanielBernoullisEpidemiological2002}%
  \BibitemOpen
  \bibfield  {author} {\bibinfo {author} {\bibfnamefont {K.}~\bibnamefont
  {Dietz}}\ and\ \bibinfo {author} {\bibfnamefont {J.~A.~P.}\ \bibnamefont
  {Heesterbeek}},\ }\bibfield  {title} {\bibinfo {title} {Daniel
  {{Bernoulli}}'s epidemiological model revisited},\ }\href
  {https://doi.org/10.1016/S0025-5564(02)00122-0} {\bibfield  {journal}
  {\bibinfo  {journal} {Mathematical Biosciences}\ }\textbf {\bibinfo {volume}
  {180}},\ \bibinfo {pages} {1} (\bibinfo {year} {2002})}\BibitemShut {NoStop}%
\bibitem [{\citenamefont {Keeling}\ and\ \citenamefont
  {Rohani}(2008)}]{keelingModelingInfectiousDiseases2008}%
  \BibitemOpen
  \bibfield  {author} {\bibinfo {author} {\bibfnamefont {M.~J.}\ \bibnamefont
  {Keeling}}\ and\ \bibinfo {author} {\bibfnamefont {P.}~\bibnamefont
  {Rohani}},\ }\href {https://doi.org/10.2307/j.ctvcm4gk0} {\emph {\bibinfo
  {title} {Modeling {{Infectious Diseases}} in {{Humans}} and {{Animals}}}}}\
  (\bibinfo  {publisher} {Princeton University Press},\ \bibinfo {year}
  {2008})\ \Eprint {https://arxiv.org/abs/j.ctvcm4gk0} {j.ctvcm4gk0}
  \BibitemShut {NoStop}%
\bibitem [{\citenamefont {Albert}\ \emph {et~al.}(2000)\citenamefont {Albert},
  \citenamefont {Jeong},\ and\ \citenamefont
  {Barab{\'a}si}}]{albertErrorAttackTolerance2000}%
  \BibitemOpen
  \bibfield  {author} {\bibinfo {author} {\bibfnamefont {R.}~\bibnamefont
  {Albert}}, \bibinfo {author} {\bibfnamefont {H.}~\bibnamefont {Jeong}},\ and\
  \bibinfo {author} {\bibfnamefont {A.-L.}\ \bibnamefont {Barab{\'a}si}},\
  }\bibfield  {title} {\bibinfo {title} {Error and attack tolerance of complex
  networks},\ }\href {https://doi.org/10.1038/35019019} {\bibfield  {journal}
  {\bibinfo  {journal} {Nature}\ }\textbf {\bibinfo {volume} {406}},\ \bibinfo
  {pages} {378} (\bibinfo {year} {2000})}\BibitemShut {NoStop}%
\bibitem [{\citenamefont {Newman}(2010)}]{newmanNetworksIntroduction2010}%
  \BibitemOpen
  \bibfield  {author} {\bibinfo {author} {\bibfnamefont {M.}~\bibnamefont
  {Newman}},\ }\href
  {https://doi.org/10.1093/acprof:oso/9780199206650.001.0001} {\emph {\bibinfo
  {title} {Networks: {{An Introduction}}}}}\ (\bibinfo  {publisher} {Oxford
  University Press},\ \bibinfo {year} {2010})\BibitemShut {NoStop}%
\bibitem [{\citenamefont {{Pastor-Satorras}}\ and\ \citenamefont
  {Vespignani}(2001{\natexlab{a}})}]{pastor-satorrasEpidemicSpreadingScaleFree2001}%
  \BibitemOpen
  \bibfield  {author} {\bibinfo {author} {\bibfnamefont {R.}~\bibnamefont
  {{Pastor-Satorras}}}\ and\ \bibinfo {author} {\bibfnamefont {A.}~\bibnamefont
  {Vespignani}},\ }\bibfield  {title} {\bibinfo {title} {Epidemic {{Spreading}}
  in {{Scale-Free Networks}}},\ }\href
  {https://doi.org/10.1103/PhysRevLett.86.3200} {\bibfield  {journal} {\bibinfo
   {journal} {Physical Review Letters}\ }\textbf {\bibinfo {volume} {86}},\
  \bibinfo {pages} {3200} (\bibinfo {year} {2001}{\natexlab{a}})}\BibitemShut
  {NoStop}%
\bibitem [{\citenamefont {{Pastor-Satorras}}\ \emph {et~al.}(2015)\citenamefont
  {{Pastor-Satorras}}, \citenamefont {Castellano}, \citenamefont
  {Van~Mieghem},\ and\ \citenamefont
  {Vespignani}}]{pastor-satorrasEpidemicProcessesComplex2015a}%
  \BibitemOpen
  \bibfield  {author} {\bibinfo {author} {\bibfnamefont {R.}~\bibnamefont
  {{Pastor-Satorras}}}, \bibinfo {author} {\bibfnamefont {C.}~\bibnamefont
  {Castellano}}, \bibinfo {author} {\bibfnamefont {P.}~\bibnamefont
  {Van~Mieghem}},\ and\ \bibinfo {author} {\bibfnamefont {A.}~\bibnamefont
  {Vespignani}},\ }\bibfield  {title} {\bibinfo {title} {Epidemic processes in
  complex networks},\ }\href {https://doi.org/10.1103/RevModPhys.87.925}
  {\bibfield  {journal} {\bibinfo  {journal} {Reviews of Modern Physics}\
  }\textbf {\bibinfo {volume} {87}},\ \bibinfo {pages} {925} (\bibinfo {year}
  {2015})}\BibitemShut {NoStop}%
\bibitem [{\citenamefont {Hays}(2005)}]{haysEpidemicsPandemicsTheir2005}%
  \BibitemOpen
  \bibfield  {author} {\bibinfo {author} {\bibfnamefont {J.~N.}\ \bibnamefont
  {Hays}},\ }\href@noop {} {\emph {\bibinfo {title} {Epidemics and Pandemics:
  Their Impacts on Human History}}}\ (\bibinfo  {publisher} {ABC-CLIO},\
  \bibinfo {address} {Santa Barbara, Calif},\ \bibinfo {year}
  {2005})\BibitemShut {NoStop}%
\bibitem [{\citenamefont {Funk}\ \emph {et~al.}(2010)\citenamefont {Funk},
  \citenamefont {Gilad},\ and\ \citenamefont
  {Jansen}}]{funkEndemicDiseaseAwareness2010}%
  \BibitemOpen
  \bibfield  {author} {\bibinfo {author} {\bibfnamefont {S.}~\bibnamefont
  {Funk}}, \bibinfo {author} {\bibfnamefont {E.}~\bibnamefont {Gilad}},\ and\
  \bibinfo {author} {\bibfnamefont {V.~A.~A.}\ \bibnamefont {Jansen}},\
  }\bibfield  {title} {\bibinfo {title} {Endemic disease, awareness, and local
  behavioural response},\ }\href {https://doi.org/10.1016/j.jtbi.2010.02.032}
  {\bibfield  {journal} {\bibinfo  {journal} {Journal of Theoretical Biology}\
  }\textbf {\bibinfo {volume} {264}},\ \bibinfo {pages} {501} (\bibinfo {year}
  {2010})}\BibitemShut {NoStop}%
\bibitem [{\citenamefont {Zhan}\ \emph {et~al.}(2018)\citenamefont {Zhan},
  \citenamefont {Liu}, \citenamefont {Zhou}, \citenamefont {Zhang},
  \citenamefont {Sun}, \citenamefont {Zhu},\ and\ \citenamefont
  {Jin}}]{zhanCouplingDynamicsEpidemic2018}%
  \BibitemOpen
  \bibfield  {author} {\bibinfo {author} {\bibfnamefont {X.-X.}\ \bibnamefont
  {Zhan}}, \bibinfo {author} {\bibfnamefont {C.}~\bibnamefont {Liu}}, \bibinfo
  {author} {\bibfnamefont {G.}~\bibnamefont {Zhou}}, \bibinfo {author}
  {\bibfnamefont {Z.-K.}\ \bibnamefont {Zhang}}, \bibinfo {author}
  {\bibfnamefont {G.-Q.}\ \bibnamefont {Sun}}, \bibinfo {author} {\bibfnamefont
  {J.~J.}\ \bibnamefont {Zhu}},\ and\ \bibinfo {author} {\bibfnamefont
  {Z.}~\bibnamefont {Jin}},\ }\bibfield  {title} {\bibinfo {title} {Coupling
  dynamics of epidemic spreading and information diffusion on complex
  networks},\ }\href {https://doi.org/10.1016/j.amc.2018.03.050} {\bibfield
  {journal} {\bibinfo  {journal} {Applied Mathematics and Computation}\
  }\textbf {\bibinfo {volume} {332}},\ \bibinfo {pages} {437} (\bibinfo {year}
  {2018})}\BibitemShut {NoStop}%
\bibitem [{\citenamefont {Samanta}\ \emph {et~al.}(2013)\citenamefont
  {Samanta}, \citenamefont {Rana}, \citenamefont {Sharma}, \citenamefont
  {Misra},\ and\ \citenamefont {Chattopadhyay}}]{Samanta2013}%
  \BibitemOpen
  \bibfield  {author} {\bibinfo {author} {\bibfnamefont {S.}~\bibnamefont
  {Samanta}}, \bibinfo {author} {\bibfnamefont {S.}~\bibnamefont {Rana}},
  \bibinfo {author} {\bibfnamefont {A.}~\bibnamefont {Sharma}}, \bibinfo
  {author} {\bibfnamefont {A.~K.}\ \bibnamefont {Misra}},\ and\ \bibinfo
  {author} {\bibfnamefont {J.}~\bibnamefont {Chattopadhyay}},\ }\bibfield
  {title} {\bibinfo {title} {Effect of awareness programs by media on the
  epidemic outbreaks: {{A}} mathematical model},\ }\href
  {https://doi.org/10.1016/j.amc.2013.01.009} {\bibfield  {journal} {\bibinfo
  {journal} {Applied Mathematics and Computation}\ }\textbf {\bibinfo {volume}
  {219}},\ \bibinfo {pages} {6965} (\bibinfo {year} {2013})}\BibitemShut
  {NoStop}%
\bibitem [{\citenamefont {Wu}\ \emph {et~al.}(2012)\citenamefont {Wu},
  \citenamefont {Fu}, \citenamefont {Small},\ and\ \citenamefont
  {Xu}}]{Wu2012}%
  \BibitemOpen
  \bibfield  {author} {\bibinfo {author} {\bibfnamefont {Q.}~\bibnamefont
  {Wu}}, \bibinfo {author} {\bibfnamefont {X.}~\bibnamefont {Fu}}, \bibinfo
  {author} {\bibfnamefont {M.}~\bibnamefont {Small}},\ and\ \bibinfo {author}
  {\bibfnamefont {X.-J.}\ \bibnamefont {Xu}},\ }\bibfield  {title} {\bibinfo
  {title} {The impact of awareness on epidemic spreading in networks},\ }\href
  {https://doi.org/10.1063/1.3673573} {\bibfield  {journal} {\bibinfo
  {journal} {Chaos}\ }\textbf {\bibinfo {volume} {22}},\ \bibinfo {pages}
  {013101} (\bibinfo {year} {2012})}\BibitemShut {NoStop}%
\bibitem [{\citenamefont {Liu}\ \emph {et~al.}(2018)\citenamefont {Liu},
  \citenamefont {Liu},\ and\ \citenamefont {Jin}}]{Liu2018}%
  \BibitemOpen
  \bibfield  {author} {\bibinfo {author} {\bibfnamefont {G.}~\bibnamefont
  {Liu}}, \bibinfo {author} {\bibfnamefont {Z.}~\bibnamefont {Liu}},\ and\
  \bibinfo {author} {\bibfnamefont {Z.}~\bibnamefont {Jin}},\ }\bibfield
  {title} {\bibinfo {title} {Dynamics analysis of epidemic and information
  spreading in overlay networks},\ }\href
  {https://doi.org/10.1016/j.jtbi.2018.02.010} {\bibfield  {journal} {\bibinfo
  {journal} {Journal of Theoretical Biology}\ }\textbf {\bibinfo {volume}
  {444}},\ \bibinfo {pages} {28} (\bibinfo {year} {2018})}\BibitemShut
  {NoStop}%
\bibitem [{\citenamefont {Zhou}\ \emph {et~al.}(2019)\citenamefont {Zhou},
  \citenamefont {Zhou}, \citenamefont {Chen},\ and\ \citenamefont
  {Stanley}}]{Zhou2019}%
  \BibitemOpen
  \bibfield  {author} {\bibinfo {author} {\bibfnamefont {Y.}~\bibnamefont
  {Zhou}}, \bibinfo {author} {\bibfnamefont {J.}~\bibnamefont {Zhou}}, \bibinfo
  {author} {\bibfnamefont {G.}~\bibnamefont {Chen}},\ and\ \bibinfo {author}
  {\bibfnamefont {H.~E.}\ \bibnamefont {Stanley}},\ }\bibfield  {title}
  {\bibinfo {title} {Effective degree theory for awareness and epidemic
  spreading on multiplex networks},\ }\href
  {https://doi.org/10.1088/1367-2630/ab0458} {\bibfield  {journal} {\bibinfo
  {journal} {New Journal of Physics}\ }\textbf {\bibinfo {volume} {21}},\
  \bibinfo {pages} {035002} (\bibinfo {year} {2019})}\BibitemShut {NoStop}%
\bibitem [{\citenamefont {Granell}\ \emph {et~al.}(2013)\citenamefont
  {Granell}, \citenamefont {G{\'o}mez},\ and\ \citenamefont
  {Arenas}}]{granellDynamicalInterplayAwareness2013}%
  \BibitemOpen
  \bibfield  {author} {\bibinfo {author} {\bibfnamefont {C.}~\bibnamefont
  {Granell}}, \bibinfo {author} {\bibfnamefont {S.}~\bibnamefont {G{\'o}mez}},\
  and\ \bibinfo {author} {\bibfnamefont {A.}~\bibnamefont {Arenas}},\
  }\bibfield  {title} {\bibinfo {title} {Dynamical {{Interplay}} between
  {{Awareness}} and {{Epidemic Spreading}} in {{Multiplex Networks}}},\ }\href
  {https://doi.org/10.1103/PhysRevLett.111.128701} {\bibfield  {journal}
  {\bibinfo  {journal} {Physical Review Letters}\ }\textbf {\bibinfo {volume}
  {111}},\ \bibinfo {pages} {128701} (\bibinfo {year} {2013})}\BibitemShut
  {NoStop}%
\bibitem [{\citenamefont {Granell}\ \emph {et~al.}(2014)\citenamefont
  {Granell}, \citenamefont {G{\'o}mez},\ and\ \citenamefont
  {Arenas}}]{Granell2014}%
  \BibitemOpen
  \bibfield  {author} {\bibinfo {author} {\bibfnamefont {C.}~\bibnamefont
  {Granell}}, \bibinfo {author} {\bibfnamefont {S.}~\bibnamefont {G{\'o}mez}},\
  and\ \bibinfo {author} {\bibfnamefont {A.}~\bibnamefont {Arenas}},\
  }\bibfield  {title} {\bibinfo {title} {Competing spreading processes on
  multiplex networks: {{Awareness}} and epidemics},\ }\href
  {https://doi.org/10.1103/PhysRevE.90.012808} {\bibfield  {journal} {\bibinfo
  {journal} {Physical Review E}\ }\textbf {\bibinfo {volume} {90}},\ \bibinfo
  {pages} {012808} (\bibinfo {year} {2014})}\BibitemShut {NoStop}%
\bibitem [{\citenamefont {Huang}\ \emph {et~al.}(2018)\citenamefont {Huang},
  \citenamefont {Juang}, \citenamefont {Liang},\ and\ \citenamefont
  {Wang}}]{Huang2018}%
  \BibitemOpen
  \bibfield  {author} {\bibinfo {author} {\bibfnamefont {Y.-J.}\ \bibnamefont
  {Huang}}, \bibinfo {author} {\bibfnamefont {J.}~\bibnamefont {Juang}},
  \bibinfo {author} {\bibfnamefont {Y.-H.}\ \bibnamefont {Liang}},\ and\
  \bibinfo {author} {\bibfnamefont {H.-Y.}\ \bibnamefont {Wang}},\ }\bibfield
  {title} {\bibinfo {title} {Global stability for epidemic models on multiplex
  networks},\ }\href {https://doi.org/10.1007/s00285-017-1179-5} {\bibfield
  {journal} {\bibinfo  {journal} {Journal of Mathematical Biology}\ }\textbf
  {\bibinfo {volume} {76}},\ \bibinfo {pages} {1339} (\bibinfo {year}
  {2018})}\BibitemShut {NoStop}%
\bibitem [{\citenamefont {Silva}\ \emph {et~al.}(2024)\citenamefont {Silva},
  \citenamefont {Rodrigues},\ and\ \citenamefont {Ferreira}}]{Silva2024}%
  \BibitemOpen
  \bibfield  {author} {\bibinfo {author} {\bibfnamefont {D.~H.}\ \bibnamefont
  {Silva}}, \bibinfo {author} {\bibfnamefont {F.~A.}\ \bibnamefont
  {Rodrigues}},\ and\ \bibinfo {author} {\bibfnamefont {S.~C.}\ \bibnamefont
  {Ferreira}},\ }\bibfield  {title} {\bibinfo {title} {Accuracy of discrete-
  and continuous-time mean-field theories for epidemic processes on complex
  networks},\ }\href {https://doi.org/10.1103/PhysRevE.110.014302} {\bibfield
  {journal} {\bibinfo  {journal} {Physical Review E}\ }\textbf {\bibinfo
  {volume} {110}},\ \bibinfo {pages} {014302} (\bibinfo {year}
  {2024})}\BibitemShut {NoStop}%
\bibitem [{\citenamefont {Pan}\ and\ \citenamefont {Yan}(2018)}]{Pan2018}%
  \BibitemOpen
  \bibfield  {author} {\bibinfo {author} {\bibfnamefont {Y.}~\bibnamefont
  {Pan}}\ and\ \bibinfo {author} {\bibfnamefont {Z.}~\bibnamefont {Yan}},\
  }\bibfield  {title} {\bibinfo {title} {The impact of multiple information on
  coupled awareness-epidemic dynamics in multiplex networks},\ }\href
  {https://doi.org/10.1016/j.physa.2017.08.082} {\bibfield  {journal} {\bibinfo
   {journal} {Physica A: Statistical Mechanics and its Applications}\ }\textbf
  {\bibinfo {volume} {491}},\ \bibinfo {pages} {45} (\bibinfo {year}
  {2018})}\BibitemShut {NoStop}%
\bibitem [{\citenamefont {Zheng}\ \emph {et~al.}(2018)\citenamefont {Zheng},
  \citenamefont {Xia}, \citenamefont {Guo},\ and\ \citenamefont
  {Dehmer}}]{Zheng2018}%
  \BibitemOpen
  \bibfield  {author} {\bibinfo {author} {\bibfnamefont {C.}~\bibnamefont
  {Zheng}}, \bibinfo {author} {\bibfnamefont {C.}~\bibnamefont {Xia}}, \bibinfo
  {author} {\bibfnamefont {Q.}~\bibnamefont {Guo}},\ and\ \bibinfo {author}
  {\bibfnamefont {M.}~\bibnamefont {Dehmer}},\ }\bibfield  {title} {\bibinfo
  {title} {Interplay between {{SIR-based}} disease spreading and awareness
  diffusion on multiplex networks},\ }\href
  {https://doi.org/10.1016/j.jpdc.2018.01.001} {\bibfield  {journal} {\bibinfo
  {journal} {Journal of Parallel and Distributed Computing}\ }\textbf {\bibinfo
  {volume} {115}},\ \bibinfo {pages} {20} (\bibinfo {year} {2018})}\BibitemShut
  {NoStop}%
\bibitem [{\citenamefont {Xia}\ \emph {et~al.}(2019)\citenamefont {Xia},
  \citenamefont {Wang}, \citenamefont {Zheng}, \citenamefont {Guo},
  \citenamefont {Shi}, \citenamefont {Dehmer},\ and\ \citenamefont
  {Chen}}]{Xia2019}%
  \BibitemOpen
  \bibfield  {author} {\bibinfo {author} {\bibfnamefont {C.}~\bibnamefont
  {Xia}}, \bibinfo {author} {\bibfnamefont {Z.}~\bibnamefont {Wang}}, \bibinfo
  {author} {\bibfnamefont {C.}~\bibnamefont {Zheng}}, \bibinfo {author}
  {\bibfnamefont {Q.}~\bibnamefont {Guo}}, \bibinfo {author} {\bibfnamefont
  {Y.}~\bibnamefont {Shi}}, \bibinfo {author} {\bibfnamefont {M.}~\bibnamefont
  {Dehmer}},\ and\ \bibinfo {author} {\bibfnamefont {Z.}~\bibnamefont {Chen}},\
  }\bibfield  {title} {\bibinfo {title} {A new coupled disease-awareness
  spreading model with mass media on multiplex networks},\ }\href
  {https://doi.org/10.1016/j.ins.2018.08.050} {\bibfield  {journal} {\bibinfo
  {journal} {Information Sciences}\ }\textbf {\bibinfo {volume} {471}},\
  \bibinfo {pages} {185} (\bibinfo {year} {2019})}\BibitemShut {NoStop}%
\bibitem [{\citenamefont {Wang}\ \emph {et~al.}(2019)\citenamefont {Wang},
  \citenamefont {Guo}, \citenamefont {Sun},\ and\ \citenamefont
  {Xia}}]{Wang2019}%
  \BibitemOpen
  \bibfield  {author} {\bibinfo {author} {\bibfnamefont {Z.}~\bibnamefont
  {Wang}}, \bibinfo {author} {\bibfnamefont {Q.}~\bibnamefont {Guo}}, \bibinfo
  {author} {\bibfnamefont {S.}~\bibnamefont {Sun}},\ and\ \bibinfo {author}
  {\bibfnamefont {C.}~\bibnamefont {Xia}},\ }\bibfield  {title} {\bibinfo
  {title} {The impact of awareness diffusion on {{SIR-like}} epidemics in
  multiplex networks},\ }\href {https://doi.org/10.1016/j.amc.2018.12.045}
  {\bibfield  {journal} {\bibinfo  {journal} {Applied Mathematics and
  Computation}\ }\textbf {\bibinfo {volume} {349}},\ \bibinfo {pages} {134}
  (\bibinfo {year} {2019})}\BibitemShut {NoStop}%
\bibitem [{\citenamefont {Ma}\ \emph {et~al.}(2022)\citenamefont {Ma},
  \citenamefont {Zhang}, \citenamefont {Zhao},\ and\ \citenamefont
  {Xue}}]{Ma2022}%
  \BibitemOpen
  \bibfield  {author} {\bibinfo {author} {\bibfnamefont {W.}~\bibnamefont
  {Ma}}, \bibinfo {author} {\bibfnamefont {P.}~\bibnamefont {Zhang}}, \bibinfo
  {author} {\bibfnamefont {X.}~\bibnamefont {Zhao}},\ and\ \bibinfo {author}
  {\bibfnamefont {L.}~\bibnamefont {Xue}},\ }\bibfield  {title} {\bibinfo
  {title} {The coupled dynamics of information dissemination and {{SEIR-based}}
  epidemic spreading in multiplex networks},\ }\href
  {https://doi.org/10.1016/j.physa.2021.126558} {\bibfield  {journal} {\bibinfo
   {journal} {Physica A: Statistical Mechanics and its Applications}\ }\textbf
  {\bibinfo {volume} {588}},\ \bibinfo {pages} {126558} (\bibinfo {year}
  {2022})}\BibitemShut {NoStop}%
\bibitem [{\citenamefont {Zhu}\ \emph {et~al.}(2019)\citenamefont {Zhu},
  \citenamefont {Wang}, \citenamefont {Li}, \citenamefont {Guo},\ and\
  \citenamefont {Wang}}]{Zhu2019}%
  \BibitemOpen
  \bibfield  {author} {\bibinfo {author} {\bibfnamefont {P.}~\bibnamefont
  {Zhu}}, \bibinfo {author} {\bibfnamefont {X.}~\bibnamefont {Wang}}, \bibinfo
  {author} {\bibfnamefont {S.}~\bibnamefont {Li}}, \bibinfo {author}
  {\bibfnamefont {Y.}~\bibnamefont {Guo}},\ and\ \bibinfo {author}
  {\bibfnamefont {Z.}~\bibnamefont {Wang}},\ }\bibfield  {title} {\bibinfo
  {title} {Investigation of epidemic spreading process on multiplex networks by
  incorporating fatal properties},\ }\href
  {https://doi.org/10.1016/j.amc.2019.02.049} {\bibfield  {journal} {\bibinfo
  {journal} {Applied Mathematics and Computation}\ }\textbf {\bibinfo {volume}
  {359}},\ \bibinfo {pages} {512} (\bibinfo {year} {2019})}\BibitemShut
  {NoStop}%
\bibitem [{\citenamefont {He}\ and\ \citenamefont {Zhu}(2021)}]{He2021}%
  \BibitemOpen
  \bibfield  {author} {\bibinfo {author} {\bibfnamefont {L.}~\bibnamefont
  {He}}\ and\ \bibinfo {author} {\bibfnamefont {L.}~\bibnamefont {Zhu}},\
  }\bibfield  {title} {\bibinfo {title} {Modeling the {{COVID-19}} epidemic and
  awareness diffusion on multiplex networks},\ }\href
  {https://doi.org/10.1088/1572-9494/abd84a} {\bibfield  {journal} {\bibinfo
  {journal} {Communications in Theoretical Physics}\ }\textbf {\bibinfo
  {volume} {73}},\ \bibinfo {pages} {035002} (\bibinfo {year}
  {2021})}\BibitemShut {NoStop}%
\bibitem [{\citenamefont {Yu}\ and\ \citenamefont
  {Huo}(2025{\natexlab{a}})}]{Yu2025}%
  \BibitemOpen
  \bibfield  {author} {\bibinfo {author} {\bibfnamefont {Y.}~\bibnamefont
  {Yu}}\ and\ \bibinfo {author} {\bibfnamefont {L.}~\bibnamefont {Huo}},\
  }\bibfield  {title} {\bibinfo {title} {A coupled {{UAU-DKD-SIQS}} model
  considering partial and complete mapping relationship in time-varying
  multiplex networks},\ }\href {https://doi.org/10.1016/j.eswa.2025.126887}
  {\bibfield  {journal} {\bibinfo  {journal} {Expert Systems with
  Applications}\ }\textbf {\bibinfo {volume} {273}},\ \bibinfo {pages} {126887}
  (\bibinfo {year} {2025}{\natexlab{a}})}\BibitemShut {NoStop}%
\bibitem [{\citenamefont {Yu}\ and\ \citenamefont
  {Huo}(2025{\natexlab{b}})}]{Yu2025a}%
  \BibitemOpen
  \bibfield  {author} {\bibinfo {author} {\bibfnamefont {Y.}~\bibnamefont
  {Yu}}\ and\ \bibinfo {author} {\bibfnamefont {L.}~\bibnamefont {Huo}},\
  }\bibfield  {title} {\bibinfo {title} {Effects of official information
  diffusion and rumor-related behavior adoption on epidemic transmission in
  multiplex networks},\ }\href {https://doi.org/10.1016/j.ins.2024.121414}
  {\bibfield  {journal} {\bibinfo  {journal} {Information Sciences}\ }\textbf
  {\bibinfo {volume} {689}},\ \bibinfo {pages} {121414} (\bibinfo {year}
  {2025}{\natexlab{b}})}\BibitemShut {NoStop}%
\bibitem [{\citenamefont {Chen}\ \emph {et~al.}(2023)\citenamefont {Chen},
  \citenamefont {Hu},\ and\ \citenamefont {Cao}}]{Chen2023}%
  \BibitemOpen
  \bibfield  {author} {\bibinfo {author} {\bibfnamefont {J.}~\bibnamefont
  {Chen}}, \bibinfo {author} {\bibfnamefont {M.}~\bibnamefont {Hu}},\ and\
  \bibinfo {author} {\bibfnamefont {J.}~\bibnamefont {Cao}},\ }\bibfield
  {title} {\bibinfo {title} {Dynamics of
  information-awareness-epidemic-activity coevolution in multiplex networks},\
  }\href {https://doi.org/10.1103/PhysRevResearch.5.033065} {\bibfield
  {journal} {\bibinfo  {journal} {Physical Review Research}\ }\textbf {\bibinfo
  {volume} {5}},\ \bibinfo {pages} {033065} (\bibinfo {year}
  {2023})}\BibitemShut {NoStop}%
\bibitem [{\citenamefont {Chen}\ \emph {et~al.}(2025)\citenamefont {Chen},
  \citenamefont {Zhang}, \citenamefont {Hu}, \citenamefont {Li},\ and\
  \citenamefont {Chen}}]{Chen2025}%
  \BibitemOpen
  \bibfield  {author} {\bibinfo {author} {\bibfnamefont {J.}~\bibnamefont
  {Chen}}, \bibinfo {author} {\bibfnamefont {Y.}~\bibnamefont {Zhang}},
  \bibinfo {author} {\bibfnamefont {M.}~\bibnamefont {Hu}}, \bibinfo {author}
  {\bibfnamefont {M.}~\bibnamefont {Li}},\ and\ \bibinfo {author}
  {\bibfnamefont {F.}~\bibnamefont {Chen}},\ }\bibfield  {title} {\bibinfo
  {title} {Nontrivial epidemic dynamics induced by information-driven
  awareness-activity-resource coevolution},\ }\href
  {https://doi.org/10.1103/PhysRevE.111.044301} {\bibfield  {journal} {\bibinfo
   {journal} {Physical Review E}\ }\textbf {\bibinfo {volume} {111}},\ \bibinfo
  {pages} {044301} (\bibinfo {year} {2025})}\BibitemShut {NoStop}%
\bibitem [{\citenamefont {Li}\ \emph {et~al.}(2024)\citenamefont {Li},
  \citenamefont {Dong}, \citenamefont {Zhu},\ and\ \citenamefont
  {Tian}}]{Li2024}%
  \BibitemOpen
  \bibfield  {author} {\bibinfo {author} {\bibfnamefont {L.}~\bibnamefont
  {Li}}, \bibinfo {author} {\bibfnamefont {G.}~\bibnamefont {Dong}}, \bibinfo
  {author} {\bibfnamefont {H.}~\bibnamefont {Zhu}},\ and\ \bibinfo {author}
  {\bibfnamefont {L.}~\bibnamefont {Tian}},\ }\bibfield  {title} {\bibinfo
  {title} {Impact of multiple doses of vaccination on epidemiological spread in
  multiple networks},\ }\href {https://doi.org/10.1016/j.amc.2024.128617}
  {\bibfield  {journal} {\bibinfo  {journal} {Applied Mathematics and
  Computation}\ }\textbf {\bibinfo {volume} {472}},\ \bibinfo {pages} {128617}
  (\bibinfo {year} {2024})}\BibitemShut {NoStop}%
\bibitem [{\citenamefont {Van Den~Driessche}\ and\ \citenamefont
  {Watmough}(2002)}]{VanDenDriessche2002}%
  \BibitemOpen
  \bibfield  {author} {\bibinfo {author} {\bibfnamefont {P.}~\bibnamefont {Van
  Den~Driessche}}\ and\ \bibinfo {author} {\bibfnamefont {J.}~\bibnamefont
  {Watmough}},\ }\bibfield  {title} {\bibinfo {title} {Reproduction numbers and
  sub-threshold endemic equilibria for compartmental models of disease
  transmission},\ }\href {https://doi.org/10.1016/S0025-5564(02)00108-6}
  {\bibfield  {journal} {\bibinfo  {journal} {Mathematical Biosciences}\
  }\textbf {\bibinfo {volume} {180}},\ \bibinfo {pages} {29} (\bibinfo {year}
  {2002})}\BibitemShut {NoStop}%
\bibitem [{\citenamefont {Chang}\ \emph {et~al.}(2021)\citenamefont {Chang},
  \citenamefont {Cai}, \citenamefont {Zhang},\ and\ \citenamefont
  {Wang}}]{Chang2021}%
  \BibitemOpen
  \bibfield  {author} {\bibinfo {author} {\bibfnamefont {X.}~\bibnamefont
  {Chang}}, \bibinfo {author} {\bibfnamefont {C.-R.}\ \bibnamefont {Cai}},
  \bibinfo {author} {\bibfnamefont {J.-Q.}\ \bibnamefont {Zhang}},\ and\
  \bibinfo {author} {\bibfnamefont {C.-Y.}\ \bibnamefont {Wang}},\ }\bibfield
  {title} {\bibinfo {title} {Analytical solution of epidemic threshold for
  coupled information-epidemic dynamics on multiplex networks with alterable
  heterogeneity},\ }\href {https://doi.org/10.1103/PhysRevE.104.044303}
  {\bibfield  {journal} {\bibinfo  {journal} {Physical Review E}\ }\textbf
  {\bibinfo {volume} {104}},\ \bibinfo {pages} {044303} (\bibinfo {year}
  {2021})}\BibitemShut {NoStop}%
\bibitem [{\citenamefont {{Pastor-Satorras}}\ and\ \citenamefont
  {Vespignani}(2001{\natexlab{b}})}]{PastorSatorras2001}%
  \BibitemOpen
  \bibfield  {author} {\bibinfo {author} {\bibfnamefont {R.}~\bibnamefont
  {{Pastor-Satorras}}}\ and\ \bibinfo {author} {\bibfnamefont {A.}~\bibnamefont
  {Vespignani}},\ }\bibfield  {title} {\bibinfo {title} {Epidemic dynamics and
  endemic states in complex networks},\ }\href
  {https://doi.org/10.1103/PhysRevE.63.066117} {\bibfield  {journal} {\bibinfo
  {journal} {Physical Review E}\ }\textbf {\bibinfo {volume} {63}},\ \bibinfo
  {pages} {066117} (\bibinfo {year} {2001}{\natexlab{b}})}\BibitemShut
  {NoStop}%
\bibitem [{\citenamefont {Sander}\ \emph {et~al.}(2016)\citenamefont {Sander},
  \citenamefont {Costa},\ and\ \citenamefont {Ferreira}}]{Sander2016}%
  \BibitemOpen
  \bibfield  {author} {\bibinfo {author} {\bibfnamefont {R.~S.}\ \bibnamefont
  {Sander}}, \bibinfo {author} {\bibfnamefont {G.~S.}\ \bibnamefont {Costa}},\
  and\ \bibinfo {author} {\bibfnamefont {S.~C.}\ \bibnamefont {Ferreira}},\
  }\bibfield  {title} {\bibinfo {title} {Sampling methods for the
  quasistationary regime of epidemic processes on regular and complex
  networks},\ }\href {https://doi.org/10.1103/PhysRevE.94.042308} {\bibfield
  {journal} {\bibinfo  {journal} {Physical Review E}\ }\textbf {\bibinfo
  {volume} {94}},\ \bibinfo {pages} {042308} (\bibinfo {year}
  {2016})}\BibitemShut {NoStop}%
\bibitem [{\citenamefont {Ferreira}\ \emph {et~al.}(2012)\citenamefont
  {Ferreira}, \citenamefont {Castellano},\ and\ \citenamefont
  {{Pastor-Satorras}}}]{Ferreira2012}%
  \BibitemOpen
  \bibfield  {author} {\bibinfo {author} {\bibfnamefont {S.~C.}\ \bibnamefont
  {Ferreira}}, \bibinfo {author} {\bibfnamefont {C.}~\bibnamefont
  {Castellano}},\ and\ \bibinfo {author} {\bibfnamefont {R.}~\bibnamefont
  {{Pastor-Satorras}}},\ }\bibfield  {title} {\bibinfo {title} {Epidemic
  thresholds of the susceptible-infected-susceptible model on networks: {{A}}
  comparison of numerical and theoretical results},\ }\href
  {https://doi.org/10.1103/PhysRevE.86.041125} {\bibfield  {journal} {\bibinfo
  {journal} {Physical Review E}\ }\textbf {\bibinfo {volume} {86}},\ \bibinfo
  {pages} {041125} (\bibinfo {year} {2012})}\BibitemShut {NoStop}%
\bibitem [{\citenamefont {Mata}\ and\ \citenamefont
  {Ferreira}(2015)}]{Mata2015}%
  \BibitemOpen
  \bibfield  {author} {\bibinfo {author} {\bibfnamefont {A.~S.}\ \bibnamefont
  {Mata}}\ and\ \bibinfo {author} {\bibfnamefont {S.~C.}\ \bibnamefont
  {Ferreira}},\ }\bibfield  {title} {\bibinfo {title} {Multiple transitions of
  the susceptible-infected-susceptible epidemic model on complex networks},\
  }\bibfield  {journal} {\bibinfo  {journal} {Physical Review E}\ }\textbf
  {\bibinfo {volume} {91}},\ \href {https://doi.org/10.1103/physreve.91.012816}
  {10.1103/physreve.91.012816} (\bibinfo {year} {2015})\BibitemShut {NoStop}%
\end{thebibliography}%

\end{document}